# A Survey of Kiloparsec-Scale Radio Outflows in Radio-Quiet Active Galactic Nuclei

Jack F. Gallimore, Department of Physics & Astrronomy, Bucknell University, Lewisburg, PA 17837 USA, jgallimo@bucknell.edu

David J. Axon and Christopher P. O'Dea, Department of Physics, Rochester Institute of Technology, 84 Lomb Memorial Drive, Rochester, NY 14623

Stefi A. Baum, Chester F. Carlson Center for Imaging Science, Rochester Institute of Technology, 54 Lomb Memorial Drive, Rochester, NY 14623

Alan Pedlar, The University of Manchester, Jodrell Bank Observatory, Macclesfield, Cheshire SK11 9DL

**ABSTRACT**

Seyfert galaxies commonly host compact jets spanning 10—100 pc scales, but larger structures are resolved out in long baseline, aperture synthesis surveys. Previous, targeted studies showed that kiloparsec-scale radio structures (KSRs) may be a common feature of Seyfert and LINER galaxies, and the origin of KSRs may be starburst or AGN. We report a new Very Large Array (VLA) survey of a complete sample of Seyfert and LINER galaxies. Out of all of the surveyed radio-quiet sources, we find that 44% (19 / 43) show extended radio structures at least 1 kpc in total extent that do not match the morphology of the disk or its associated star-forming regions. The detection rate is a lower limit owing to the combined effects of projection and resolution. The infrared colors of the KSR host galaxies are unremarkable compared to other Seyferts, and the large-scale outflows orient randomly with respect to the host galaxy axes. The KSR Seyferts instead stand out by deviating significantly from the far-infrared – radio correlation for star-forming galaxies, with tendency towards radio excess, and they are more likely to have a relatively luminous, compact radio source in the nucleus; these results argue that KSRs are powered by the AGN rather than starburst. The high detection rate indicates that Seyferts generate radio outflows over a significant fraction of their lifetime, which is much longer than the dynamical timescale of an AGN-powered jet but comparable instead to the buoyancy timescale. The likely explanation is that the KSRs originate from jet plasma that has been decelerated by interaction with the nuclear ISM. Based on a simple ram pressure argument, the kinetic power of the jet on kiloparsec scales is about three orders of magnitude weaker than the power of the jet on 10—100 pc scales. This result is consistent with the interaction model, in which case virtually all of the jet power must be lost to the ISM within the inner kiloparsec.

# 1 INTRODUCTION

As a class, Seyfert galaxies show all of the characteristics of more luminous active galactic nuclei (AGNs), including broad optical line emission, variability, and compact nuclear radio emission, which includes contributions both from star formation and the AGN (Wilson & Willis 1980; Condon et al. 1982). High angular resolution studies, which resolve out the emission related to star-formation, found that the compact, nuclear radio structures can be complex, comprising multiple, aligned components, linear or S-shaped features, lobes, and loops. The general consensus is that this high brightness temperature emission arises from low-power radio jets and outflows, analogous to the larger jets found in radio galaxies, but perhaps distorted or stunted by interaction with the surrounding ISM of the host spiral galaxy (Ulvestad, Wilson, & Sramek 1981; Booler, Pedlar, & Davies 1982; Wilson & Ulvestad 1982b; Neff & de Bruyn 1983; Pedlar et al. 1983; Wilson & Ulvestad 1983; Pedlar, Unger, & Dyson 1985; Wilson & Ulvestad 1987; Gallimore, Baum, & O'Dea 1996a; Gallimore et al. 1999; Whittle & Wilson 2004).

Radio imaging surveys of statistically well-defined samples of Seyfert galaxies have mainly employed long interferometric baselines (i.e., VLA A-array, or MERLIN) that filter out low surface brightness emission distributed over larger angular scales (Ulvestad & Wilson 1984b, 1984a; Ulvestad & Wilson 1989; Kukula et al. 1995; Thean et al. 2000; Thean et al. 2001b). These surveys may have therefore artificially suppressed the detection of larger scale outflows. For instance, while the typical jet length scale in these long-baseline surveys is ~ 10 pc – a few hundred pc, early, targeted VLA observations, obtained with shorter baselines than the later surveys, found larger scale radio outflows, exceeding 1 kpc in full extent (Wilson & Willis 1980; Hummel, van Gorkom, & Kotanyi 1983). One of the more spectacular examples is Mrk 6, which displays a ~ 600 pc nuclear jet roughly aligned with 10 kpc-scale lobes (Baum et al. 1993; Kukula et al. 1996). In fact, such large scale outflows may be common: Baum et al. (1993) used the Westerbork synthesis telescope (~ 5″ resolution at the $\nu$ = 5 GHz) to observe a sample of 13 radio-bright Seyfert galaxies, and, remarkably, 12 of these sources showed extended radio lobes extending up to several kpc from the nucleus.

The questions put forward by these results are whether such kiloparsec-scale radio structures (KSRs) are truly are so common and whether they are related to the AGN or a starburst. For example, the KSR might be an extension of the nuclear radio jet or perhaps trace relic jet material connected to a past episode of nuclear activity. On the other hand, KSRs are found in starburst galaxies, such as M82 (Seaquist & Odegard 1991) and NGC 253 (Carilli et al. 1992); such starburst KSRs probably originate from a superwind generated by the cumulative effects of stellar winds and supernovae (Chevalier & Clegg 1985; Heckman, Armus, & Miley 1987; McCarthy, van Breugel, & Heckman 1987; Heckman, Armus, & Miley 1990). Superwinds escape along the rotation axis of the host galaxy, and associated KSRs therefore appear oriented (in projection, at least) along the minor axis of the host galaxy. Seven of the 12 Seyfert KSR sources in the Baum et al. (1993) survey align within 30° of the galaxy minor axis. The ratios of the KSR radio flux density to the IRAS 60μm flux density furthermore fall with the narrow range of values found for star-forming spiral galaxies. The authors argued in favor of a starburst origin for those particular Seyfert KSRs.

Colbert et al. (1996a) used the VLA in C-configuration (~ 5″ angular resolution at $\nu$ = 5 GHz) to study a sample of ten edge-on Seyfert galaxies and four star-forming (non-Seyfert) galaxies reasonably matching the Seyfert sample in morphology, redshift, size, and inclination. The main result was, again, that KSRs are common among Seyfert galaxies, even more common than in starburst galaxies at the sensitivity of the survey. There are also some clear morphological differences between the Seyfert and starburst radio sources: Seyfert KSRs are commonly show confined structures resembling the radio lobes of radio galaxies, whereas the starburst radio structures appear more spherically symmetric (i.e., wind-like). Moreover, they found that the Seyfert KSRs do not necessarily align with the minor axis of the host galaxy, but the starburst KSRs always align with the minor axis. Colbert et al. concluded that the Seyfert KSRs probably result from directed outflows that arise from the nuclear jet but that may have been diverted or otherwise affected by a starburst-driven wind or interactions with ISM. The scale of these outflows also raises the possibility that they may be relics of past activity (e.g., Sanders 1984).

Towards improving our understanding of Seyfert radio structures, we present a VLA[1] search for KSRs in a statistically well-defined sample of optically-selected and infrared-selected Seyfert galaxies. In contrast to the sample of Colbert et al. (1996a), the present sample was not limited by host-galaxy inclination and is larger: 43 active galaxies in the present sample vs. 10 in Colbert et al. We further employed the VLA in D-configuration for improved surface brightness sensitivity. The primary goals of the present study are to assess the commonness of Seyfert KSRs and to search for clues as to their origin. This paper is organized as follows. The observations and data handling are discussed in Section 2. The primary results, including how KSRs are identified, the detection fraction, and details on individual objects, are presented in Section 3. Section 4 provides analysis of the detection statistics, including a comparison of KSR and non-KSR Seyferts. We discuss the implications of these statistics and relevant timescales in Section 5 and summarize the main results in Section 6.

For the purpose of assessing the linear scales of the radio sources, we adopt $H_0$ = 70 km s$^{-1}$ Mpc$^{-1}$. The exceptions are NGC 4388, NGC 4501, and NGC 4579, which belong to the Virgo cluster (Binggeli, Sandage, & Tammann 1985), and we assume a distance of 16 Mpc to these galaxies (Graham et al. 1999).

---

[1] The Very Large Array is operated by the National Radio Astronomy Observatory, which is a facility of the National Science Foundation operated under cooperative agreement by Associated Universities, Inc.

## 2 OBSERVATIONS AND DATA REDUCTION

### *2.1 Sample Selection*

43 Seyfert galaxies were selected from the CfA (Huchra & Burg 1992) and extended 12 µm (Rush, Malkan, & Spinoglio 1993) surveys of nearby AGNs. Four of these galaxies also have been classified as LINERs (galaxies with a "low ionization nuclear emission-line region," after Heckman 1980) according to the NASA/IPAC Extragalactic Database (NED); see Table 1. For the purposes of comparing type 1 and type 2 Seyferts (Section 4.5), we adopted the Seyfert classification provided by NED.

The CfA and 12 µm surveys were chosen because of the existing database of high resolution VLA, MERLIN, and VLBA observations already available for comparison (Kukula et al. 1995; Murray et al. 1999; Thean et al. 2000; Thean et al. 2001b). At $cz \sim$ 5000 km s$^{-1}$, the present observations provide a linear resolution of $\sim$ 5 kpc. For comparison, the KSR of Mrk 6 extends $\sim$ 7 kpc to either side of the nuclear (100 pc-scale) radio source. We therefore restricted our observations to the redshift limited by the Seyfert galaxy NGC 5548 ($cz \leq 5149$ km s$^{-1}$) to ensure that we could potentially resolve Mrk 6-like KSRs from bright, nuclear radio sources in these objects. The sample was further restricted to $\delta > -20°$ for access from the VLA site.

The sample comprises 20 type 1 or intermediate type 1 (1.$n$) Seyferts and 23 type 2 Seyferts. There may be a bias against distant, low luminosity type 2s in this joint CfA-12µm sample, however, since they are more difficult to identify by optical spectroscopy, and, owing perhaps to anisotropic emission, they be weaker mid-infrared sources compared to comparably luminous Seyfert 1s (Heckman 1995; Kukula et al. 1995; Maiolino et al. 1995; Thean et al. 2001a). The mid-infrared selection effect has been under debate, however (Spinoglio et al. 1995; Fadda et al. 1998; Nenkova, Ivezic, & Elitzur 2002), but Thean et al. (2001a) demonstrated that there is no evidence for selection against low luminosity Seyfert 2s relative to Seyfert 1s particularly in the 12µm sample.

To check for possible luminosity bias in the joint CfA-12µm sample, we performed the Kolmogorov-Smirnov (KS) test to compare the redshift distributions of type 1s vs. type 2s. If distant Seyfert 2s are missing relative to comparably luminous Seyfert 1s, the distributions should be statistically distinguishable with Seyfert 2s sampling a smaller range of redshift. For the purposes of the test we grouped the intermediate type 1s with the type 1 sample. The KS test found no significant difference in the redshift distribution of the type 1s and type 2s; the probability that they were drawn from the same parent distribution of redshifts is $P = 0.91$.

We also compared the distribution of 12 µm luminosities ($L_{12}$) calculated from measurements in the IRAS Faint Source Catalog (Moshir et al. 1992). Spinoglio and collaborators have argued that $L_{12}$ scales almost linearly with the bolometric luminosity of the host galaxy, including the contribution of the AGN (Spinoglio & Malkan 1989; Spinoglio et al. 1995). Again, the KS test found no significant difference in the distribution of log $L_{12}$ with $P = 0.99$. As a side result, the fact that there is no significant difference of either the redshift distribution or $L_{12}$ distribution argues that mid-infrared emission is not significantly enhanced in type 1s compared to the type 2s in the sample. We conclude that the joint CfA-12μm sample affords a fair comparison of type 1s and type 2s insofar as there is no strong bias resulting from luminosity selection effects.

Table 1 summarizes the 40 Seyfert galaxies observed for this survey. The Seyfert types and recessional velocities were obtained from the NED database. The Hubble classifications were taken either from the RC3 (de Vaucouleurs et al. 1991) or NED and were coded according to the numerical scheme of the RC3. Of the sources that remained after filtering for the redshift and declination limits, we chose not to re-observe four well-studied galaxies: M51 (Ford et al. 1985; Crane & van der Hulst 1992), NGC 3079 (de Bruyn 1977; Duric et al. 1983; Irwin & Saikia 2003), NGC 3516 (Wrobel & Heeschen 1988; Baum et al. 1993), and NGC 4151 (Baum et al. 1993; Pedlar et al. 1993). These four sources were however included in the sample in the statistical analyses to follow (see Section 4). The more distant Seyfert Mrk 9 was observed serendipitously to accommodate the observing schedule. The breakdown of the entire sample by Seyfert type is 5 type 1s, 15 intermediate type 1s, and 23 type 2s.

## 2.2 Observations and data reduction

All of the observations were made with the VLA operating in D-configuration and using the standard 5 GHz continuum mode. The observations took place on 20 Feb., 23 Feb, and 3 Mar. 2003. Each source was observed for a total integration time of at least 11 minutes with bracketing scans of a nearby phase reference source. The characteristic angular resolution (synthetic beam) for this observing configuration is 15″–20″, depending on declination (Table 1).

Data reduction was performed using standard procedures in the AIPS software environment. After flagging of obviously spurious data, the flux scale was established based on scans of 3C 48 and 3C 286 and bootstrapped to the phase references. Phases were calibrated against point source models for the phase references, and the solutions were transferred by interpolation in time to the target sources. Initial images were generated using numerical Fourier transforms and CLEAN deconvolution (using the AIPS task IMAGR), and all of the observed Seyfert galaxies were detected. The phases were subsequently self-calibrated based on CLEAN models of the Seyfert radio sources and neighboring sources within the primary beam. The typical background noise level (rms) of the final images is ~0.05 mJy beam$^{-1}$ (Table 1), or roughly 4 mK in terms of brightness temperature. Note that the brightness temperatures of spiral galaxy disks are of order tens of mK (Hummel 1980; Condon et al. 1982), and so the present observations are sensitive to disk radio emission from the host galaxy.

### 2.2.1 Point source subtraction

Most of the Seyfert galaxies observed for this survey harbor compact radio structure that is unresolved by the VLA C-array and is sufficiently bright to hide fainter, extended structure even outside the beam-width (cf. Baum et al. 1993). To uncover potentially hidden radio continuum structure, we subtracted the brightest unresolved component from each of the CLEAN images. The results are summarized in Table 2. The procedure involved simultaneously fitting three components within a fitting window, usually 4-5 beamwidths on a side. The fitted components comprise (1) a gaussian matched to the beam-size, i.e., a point source; (2) a second, extended gaussian intended to represent any underlying radio structure; and (3) a flat background to account for smooth structure larger than the fitting window. The unresolved component (component 1) was then subtracted from the image, leaving behind only the resolved emission.

### 2.2.2 Archival VLA data

The present observations of NGC 1241 revealed peculiar, external radio structure that might arise from a background radio source (in the discussion to follow, we show that the peculiar structure is probably a background radio galaxy). To investigate this source in further detail, we retrieved archival VLA observations taken at $\nu = 1.5$ GHz. The observations occurred on 12 March 1984 with the VLA in its B-configuration. The data, which were taken as part of program AD100, were intended to study the radio source B0308-090, but NGC 1241 appears within the field.

We reduced the archival data as described in Section 2.2. The flux scale was set by a scan of 3C 138, and phase corrections were generated based on scans of the nearby calibrator sources B0238-084, B0240-217, and B0336-019. The data were converted to J2000 coordinates using the AIPS task UVFIX. Subsequent imaging was performed by Fourier inversion and CLEAN deconvolution with a forced restoring beam of 7.5″ FWHM, which slightly super-resolves the image along the declination axis.

## 3 RESULTS

All but five of the observed Seyfert galaxies (and Mrk 9) showed resolved structure after subtraction of the brightest point source. Except for a few cases, however, the radio source morphology alone does not suffice to interpret the nature of the extended radio emission. Distinguishing emission from KSRs and emission related to star-formation in the galaxy disk presents the chief difficulty, particularly for the marginally resolved sources. To help resolve the ambiguity, we laid the radio continuum images over Digitized Sky Survey[2] (DSS) images. The results are displayed in Figure 1 – Figure 3.

---

[2] The Digitized Sky Survey was produced at the Space Telescope Science Institute under U.S. Government grant NAG W-2166. The images of these surveys are based on photographic data obtained using the Oschin

Based on the radio morphology and comparison with the DSS overlays, the Seyfert radio sources were broadly classified into three categories: (1) point sources, (2) disks, and (3) KSR candidates. The classifications are included in Table 2. Point sources, presented in Figure 1, are defined by a lack of significant residual emission after point source subtraction. Disk sources, displayed in Figure 2, are identified by a good morphological correspondence between the radio source and the optical galaxy; for example, a disk source might be so classified based on a match up of radio and optical spiral arms, or, for poorly resolved sources, a broad match between the shapes of the radio continuum and the brighter optical disk. Finally, the KSR candidates, Figure 3, were classified based on the presence of residual emission that meets or exceeds 1 kpc in total extent but could not be clearly matched to any structure of the optical galaxy. Sources known to have extended radio structure based on other observations but which are not resolved by the present observations were also placed in the KSR category. Several sources were difficult to distinguish owing to the faintness of the residual emission; we conservatively classified these ambiguous cases as disks if the extended radio emission fit within the bright optical disk on the DSS image. Note that these categories are not intended to describe the dominant radio source of the galaxy; if a KSR was detected at all, the source was classified as a KSR even if the disk is brighter in integrated flux density (NGC 4388 is an example).

The breakdown for the 40 Seyferts observed as part of this survey is 5 point sources, 19 disks, and 16 KSR candidates. There were three KSR sources that met the sample selection criteria but were not included in the observations, giving a total detection rate of 19 KSRs out of 43 sample Seyfert galaxies.

## 3.1 *Notes on the KSR candidates*

The radio sources from our sample that are classified as KSR candidates and their basic properties are listed in Table 3. The list includes several KSR sources that were detected at lower frequencies than the present observations, or that could not be resolved by the present observations, or which we chose not to re-observe because they were already known KSR candidates. These sources are indicated by references in parentheses. The reported KSR position angles refer specifically to extended radio structure apart from obvious disk emission. The "extent" refers to the end-to-end (or lobe-to-lobe) extent of the radio source, usually measured by image moment analysis, but not necessarily the extent with respect to the nucleus. The morphological classifications include "lobes," meaning radio peaks that clearly disconnect from the nuclear or disk radio structure; "tongues," which are extended radio sources that join continuously with the nuclear radio source; "linear," which refers to sources that appear jet-like or are otherwise narrower than the "tongue" sources, and "loops," which include figure-8, X-shaped, or other edge-brightened lobe-like structures. The "sidedness" refers to the presence of predominately one-sided or two-sided structures with respect to the nucleus. The morphology and

---

Schmidt Telescope on Palomar Mountain and the UK Schmidt Telescope. The plates were processed into the present compressed digital form with the permission of these institutions.

sidedness classification necessarily depend on the resolution and sensitivity of the observations. In every case, the morphology and sidedness refer to the appearance on the finest resolution radio images that detect the extended KSR structure.

We detail below the properties of the individual KSR radio sources.

### 3.1.1 Mrk 348 (NGC 262)

Mrk 348 is luminous Seyfert 2 galaxy (Arakelian, Dibay, & Esipov 1972; Koski 1978) that displays clear polarimetric evidence for a hidden Seyfert 1 nucleus (Miller & Goodrich 1990). The radio continuum emission is dominated by a variable, VLBI-scale (~ 0.5 pc) jet (Ulvestad et al. 1999) that feeds into a larger (~ 60 pc) linear radio structure oriented roughly north-south (Neff & de Bruyn 1983; Anton et al. 2002). Baum et al. (1993) discovered large-scale radio lobes (~ 6 kpc extent) that roughly align with the small-scale jet structure. The present observations marginally resolve the large-scale lobes after point-source subtraction.

### 3.1.2 NGC 1068

NGC 1068 is one of the classical Seyfert galaxies (Seyfert 1943) and is well-known for harboring a hidden type 1 nucleus (Antonucci & Miller 1985). The radio continuum resolves into a 3 kpc diameter, actively star-forming disk (Wynn-Williams, Becklin, & Scoville 1985; Gallimore et al. 1996c) and a kiloparsec-scale radio jet that terminates in parabolic, edge-brightened lobes which straddle the nucleus (Wilson & Ulvestad 1987; Gallimore et al. 1996c). The present observations marginally resolve the twin lobes after point-source subtraction. The central arcsecond further resolves into a 100 pc-scale jet and compact radio knots (Gallimore et al. 1996c; Muxlow et al. 1996). The compact jet appears to have been deflected by interaction with the circumnuclear ISM (Gallimore et al. 1996a; Gallimore et al. 1996b).

### 3.1.3 NGC 1241

NGC 1241 is a Seyfert 2 galaxy (Dahari 1985) interacting with the neighboring spiral galaxy NGC 1242 (Keel 1996 and references therein). Our observations (Figure 3) show extended radio continuum emission associated with the optical disk, and, peculiarly, two compact radio sources offset by ~ 40″ − 60″ northeast and east of the nucleus that resemble a single-sided ejection structure. The relative brightness and misaligned position angle of the two offset sources however suggest that they might be associated with a background source or sources. Figure 4 shows an archival HST/WFPC-2 image with the archival VLA 1.5 GHz image overlaid as contours. The HST image reveals a resolved, $12^{th}$ magnitude (in F606W) galaxy, which also appears as a faint knot on the DSS image, located between the two offset radio sources. There are also a handful of background galaxies that might be part of an associated cluster. It therefore seems likely that the double radio source to the northeast is actually a background source.

The archival 1.5 GHz image however reveals emission extending ~ 16″ (4.5 kpc) from the nucleus towards the east (PA 100°) that cannot be resolved in the present 5 GHz survey. There is no matching morphology in the surrounding galaxy disk to suggest a significant contribution owing to disk star formation, and so we classify this source as a KSR candidate.

### 3.1.4 NGC 1320

NGC 1320 is a low luminosity Seyfert 2 galaxy (de Robertis & Osterbrock 1986). Colbert et al. (1996a) first reported the faint radio continuum structure extending 5″ (~ 1 kpc) south of the nucleus. The present observations reveal still fainter radio emission continuing to ~ 14″ (2.6 kpc) south. Longer baseline measurements resolve out the KSR, but the nuclear radio continuum source remains unresolved with size < 0.2″ (< 40 pc) (Thean et al. 2000). An HST/WFPC-2 image in the light of [OIII] $\lambda$5007 reveals emission extended ~ 2″ (370 pc) to the northwest along the galaxy major axis but misaligned with the faint radio extension (Ferruit, Wilson, & Mulchaey 2000).

### 3.1.5 NGC 2639

NGC 2639 has been classified as a type 1 Seyfert, owing to a broad component of H$\alpha$ (Huchra, Wyatt, & Davis 1982; Keel 1983), or LINER (Ho, Filippenko, & Sargent 1993), and there is a compact, H$_2$O megamaser source argued to trace a molecular accretion disk (Wilson, Braatz, & Henkel 1995; Wilson et al. 1998 and references therein). The nucleus also contains a linear, jet-like radio source, 1.6″ (370 pc) in extent (Ulvestad & Wilson 1984b; Gallimore et al. 1999; Thean et al. 2000). The AGN is marked by a ~ 7 mas (1.6 pc), VLBI-scale radio jet (Hummel et al. 1982; Wilson et al. 1998) that aligns with the arcsecond-scale jet. The present observations show additional radio emission extending ~53″ (~ 12 kpc) and roughly aligned with the compact radio source axis.

### 3.1.6 NGC 2992

NGC 2992 is an edge-on, type 2 narrow line X-ray galaxy (Ward et al. 1978; Ward et al. 1980; Huchra et al. 1982) with an obscured BLR (e.g., Goodrich, Veilleux, & Hill 1994). Evidence for a large scale outflow, closely aligned with the galaxy minor axis, comes from observations of extended optical emission line gas (e.g., Veilleux, Shopbell, & Miller 2001) and radio continuum (Ward et al. 1980; Hummel et al. 1983). Detected at 1.4 GHz, there is a plume of radio continuum extending up to ~ 81″ (12.5 kpc) east of the radio nucleus. Our present observations fail to detect this plume at 5 GHz (cf. Colbert et al. 1996b). The extended radio source is therefore probably steep-spectrum and well-resolved by the VLA D-array beam; for comparison, the beam area of our observations is about 10% that of the Fleurs synthesis array employed by Ward et al. (1980). The nucleus further resolves into a smaller, bipolar, "figure-8" radio structure that spans ~ 9″ (1.4 kpc) across a compact nuclear source (Wehrle & Morris 1988; Thean et al. 2000). The "figure-8" radio structure is oriented at nearly right angles with respect to the larger scale radio axis.

### 3.1.7 NGC 3079

NGC 3079 is an edge-on spiral galaxy, variously classified as a Seyfert (Iyomoto et al. 2001; Irwin & Saikia 2003) or LINER (Heckman 1980; Keel 1983), the controversy arising from the interpretation of a broad H$\alpha$ emission component (Ho, Filippenko, & Sargent 1997b). It is also a famous "superwind" galaxy, evidenced by optical emission line gas, "figure-8" radio lobes, and an X-ray emitting cone, all aligned and extended along the galaxy minor axis (Duric & Seaquist 1988; Cecil et al. 2001; Cecil, Bland-Hawthorn, & Veilleux 2002; Irwin & Saikia 2003). The radio lobes span ~ 45″ (3.6 kpc) (Duric et al. 1983; Duric & Seaquist 1988; Gallimore et al. 1994), and low frequency ($\nu <$ 1.4 GHz) observations reveal radio continuum emission out to 11 kpc from the nucleus (Irwin & Saikia 2003). The nucleus harbors a VLBI-scale radio jet rotated by $\Delta$PA ~ 60° from the large scale, minor axis outflow and a compact H$_2$O megamaser source (Irwin & Seaquist 1988; Trotter et al. 1998)

### 3.1.8 NGC 3516

Long-baseline VLA observations of this classical Seyfert 1 nucleus (Seyfert 1943; Khachikian & Weedman 1974) show a compact, flat spectrum core with a 0.7″ (120 pc), one-sided northern extension along PA 8° (Nagar et al. 1999). At arcsecond resolution, the VLA detects additional components extending out to 4″ (740 pc) with PA rotating to ~ 20° and aligned with a cospatial, S-shaped optical emission line structure (Miyaji, Wilson, & Perez-Fournon 1992; Ferruit, Wilson, & Mulchaey 1998). There appears at still coarser resolutions a linear radio structure spanning ~ 45″ (8.3 kpc), oriented along PA 44°, across the nucleus (Wrobel & Heeschen 1988; Baum et al. 1993).

### 3.1.9 NGC 4051

NGC 4051 contains a low-luminosity Seyfert 1 nucleus (Khachikian & Weedman 1974; Huchra & Burg 1992; Ho et al. 1997b) and a compact H$_2$O megamaser source (Hagiwara et al. 2003). Observed by the VLA at 5 GHz, the nucleus contains a 0.4″ (20 pc) double radio source oriented along PA 81° (Ulvestad & Wilson 1984b), and 1.4 GHz observations reveal a ~ 2″ (100 pc) lobe extending southwest of the nucleus along PA ~ 145° (Ulvestad & Wilson 1984b). Baum et al. (1993) mapped radio continuum emission out to 15″ (1.5 kpc) from the nucleus along PA 32° (i.e., in the direction opposite the 2″ southwestern lobe). The large-scale structure was described as banana-shaped, and the present observations marginally resolve this structure after subtraction of the nuclear point source. We also detect emission associated with the spiral arm structure of the galaxy disk (cf. Baum et al. 1993).

### 3.1.10 NGC 4151

NGC 4151 is a classical Seyfert galaxy (Seyfert 1943) with a type 1 optical spectrum (Khachikian & Weedman 1974). The nuclear radio source resolves into compact knots aligned in a slightly curved jet structure spanning ~ 5″ (340 pc) across the nucleus and oriented along PA 80° (Ulvestad et al. 1981; Booler et al. 1982; Wilson & Ulvestad 1982b; Pedlar et al. 1993; Mundell et al. 2003). The WSRT maps of Baum et al. (1993) reveal a ~ 15″ (1 kpc) linear structure, roughly aligned with the inner jet, and a bracketing pair of radio continuum arcs located ~ 45″ (3 kpc) from the radio nucleus.

### 3.1.11 NGC 4235

NGC 4235 is a highly inclined Seyfert 1 galaxy (e.g., Keel 1980). The central radio source is unresolved by long-baseline VLA and MERLIN observations (Ulvestad & Wilson 1984b; Kukula et al. 1995; Ho & Ulvestad 2001; Thean et al. 2001b). Colbert et al. (1996a) first detected the arcminute-scale radio features located east and west of the nucleus in their 5 GHz, VLA C-array images. The present D-array images recover additional, extended emission, and the radio structures resemble lobes of a jet connecting back to the central, unresolved continuum source.

### 3.1.12 NGC 4388

NGC 4388 is a highly inclined Seyfert 2 galaxy located in the Virgo cluster. It harbors a hidden BLR visible only in polarized light (Young et al. 1996). The arcsecond-scale radio structure comprises a nuclear 2″ (150 pc) double source, oriented along PA ~ 20°, and a fainter lobe centered ~ 12″ (950 pc) north of the nucleus (Stone, Wilson, & Ward 1988). The NLR is extended along the radio source axis and appears conical in shape (Pogge 1988; Falcke, Wilson, & Simpson 1998). Plumes of radio emission continue ~ 20″ (1.6 kpc) to the north and south, roughly aligned with the minor axis of the host galaxy (Hummel et al. 1983; Stone et al. 1988), which the present observations marginally resolve from disk radio continuum emission. There is also evidence for kiloparsec-scale, minor axis outflow based on a study of optical emission line kinematics (Veilleux et al. 1999) and the presence of extended X-ray emission (Matt et al. 1994).

### 3.1.13 NGC 4593

NGC 4593 contains a type 1 Seyfert nucleus (Bell Burnell & Culhane 1979; MacAlpine, Williams, & Lewis 1979). Arcsecond and subarcsecond resolution radio observations detect only a point source (Ulvestad & Wilson 1984b; Schmitt et al. 2001; Thean et al. 2001a), but the present observations newly reveal faint radio lobes, 26″ (4.9 kpc) in length, along PA ~ 90°. There is also a plume of [OIII] $\lambda 5007$ emission extending ~ 1.7″ (320 pc) along this axis (Schmitt et al. 2003).

### 3.1.14 NGC 4594

NGC 4594, the famous "Sombrero" galaxy, contains a LINER 2 nucleus (Heckman 1980; Ho et al. 1997b) and bright (~ 100 mJy, variable), unresolved radio continuum source (Hummel, van der Hulst, & Dickey 1984; Condon 1987; Bajaja et al. 1988). The present observations reveal a fainter, 54″ (3.8 kpc) linear radio structure, with still fainter emission spanning ~ 95″ (6.7 kpc), oriented obliquely with the galaxy minor axis. This structure has never before been seen on total intensity (Stokes *I*) images, probably because it is too faint on long baselines, or because of confusion with the nuclear point source on short baselines. However, Bajaja et al. (1988) found structure similar in orientation and extent on VLA images of linearly polarized emission at 1.4 GHz. It seems likely that this faint, but highly polarized, linear radio feature is a weakly emitting KSR.

### 3.1.15 NGC 5347

NGC 5347 is a Seyfert 2 galaxy with a weak, hidden BLR (Moran et al. 2001). Arcsecond and sub-arcsecond resolution VLA observations find only an unresolved nuclear radio source (Ulvestad & Wilson 1989; Schmitt et al. 2001; Thean et al. 2001a). The present observations recover radio continuum extending ~ 16″ (2.6 kpc) south of the nucleus. In contrast, the resolved NLR extends ~ 2″ (325 pc) to the northeast (Pogge 1989; Gonzalez Delgado & Perez 1996; Schmitt et al. 2003).

### 3.1.16 NGC 5506

NGC 5506 has been classified as a narrow-line X-ray galaxy (Wilson et al. 1976), Seyfert 2 galaxy (Rubin 1978), and an "almost Seyfert" galaxy (Glass 1978). Nagar et al. (2002) recently classified it as a narrow-line Seyfert 1 galaxy with evidence for a hidden BLR based on near-infrared spectroscopy (cf. Veilleux, Goodrich, & Hill 1997). NGC 5506 is also an $H_2O$ megamaser source (Braatz, Wilson, & Henkel 1994). The radio structure at arcsecond and sub-arcsecond resolution comprises an unresolved nuclear source surrounded by diffuse emission (Ulvestad et al. 1981; Unger et al. 1986). The diffuse emission resolves into a ~ 2″ (240 pc) diameter loop centered ~ 1″ (120 pc) north (i.e., along the galaxy minor axis) of the unresolved nuclear source, and fainter extensions appear to the south (Wehrle & Morris 1987). Colbert et al. (1996a) discovered extended, amorphous radio emission spanning ~ 45″ (5.9 kpc) across the nucleus along PA 140°. The present observations marginally resolve the larger radio structure. Soft X-ray emission also extends ~ 1 – 2″ (120 – 240 pc) along the radio axis (Bianchi et al. 2003).

### 3.1.17 NGC 5548

NGC 5548 is a well-studied Seyfert 1 galaxy (Weedman 1972) with variable optical / UV continuum and broad emission lines (e.g., Peterson 1993 and references therein). The high resolution VLA images of Wilson & Ulvestad (1982a) reveal a compact core between two diffuse lobes separated by ~ 15″. The present observations marginally resolve the lobes after subtraction of the nuclear point source. There are also two radio sources lying ~ 50″ and ~ 100″ west of the nuclear radio source that correspond to sources "3" and "4" of Wilson & Ulvestad. The rough alignment of the sources suggests an origin in NGC 5548, but we cannot rule out the possibility that these radio sources are background objects. No counterparts are apparent either on the DSS image (cf. Wilson & Ulvestad) or on 2MASS images (point source limiting magnitude $K_s$ < 14.3). We also searched an archival HST/WFPC-2 image that overlapped with source 3 (the middle radio source), and again no optical counterpart was found to a limiting magnitude $m_{606W}$ ~ 21 near the position of source 3. The field-of-view did not cover enough area between sources 3 & 4 to search for a counterpart there.

### 3.1.18 NGC 5695

NGC 5695 (Mrk 686) is a type 2 Seyfert galaxy (Afanasiev et al. 1980). At ~ arcsecond resolution, the radio structure comprises only a steep-spectrum, unresolved nuclear source (Ulvestad & Wilson 1989; Kukula et al. 1995). The present observations reveal a faint, one-sided structure stretching ~ 18″ (5.3 kpc) from the nucleus and aligned within ~ 10° of the galaxy minor axis.

### 3.1.19 NGC 5929

NGC 5929 contains a type 2 Seyfert nucleus (e.g., Huchra et al. 1982). The nuclear radio structure resolves into two steep spectrum components separated by 1.3″ (220 pc) along PA 70° and bracketing a central, flat spectrum source (Ulvestad & Wilson 1984b; Wilson & Keel 1989; Su et al. 1996). The NLR extends along the radio axis, and there are emission line clouds associated with the steep spectrum radio source (Whittle et al. 1986; Taylor et al. 1989; Wilson & Keel 1989; Ferruit et al. 1999). Baum et al. (1993) discovered radio emission spanning 15″ (2.6 kpc) across the nucleus along PA 172°.

### 3.1.20 M 51

M51 ("the Whirlpool Galaxy") is the well-known interacting pair of galaxies NGC 5194 and NGC 5195; we mention it here as a notable also-ran. Both nuclei are classified as LINERs (Heckman 1980; Huchra & Burg 1992) The more prominent spiral galaxy, NGC 5194, harbors a remarkable compact radio jet that appears to fill edge-brightened radio lobes straddling the nucleus (Ford et al. 1985; Crane & van der Hulst 1992). The projected extent of the nuclear radio source is ~ 800 pc, which falls below the 1 kpc cutoff of the present survey.

# 4 ANALYSIS

## 4.1 *Intrinsic Properties of the KSR Sources*

We estimated the intrinsic properties of the KSR radio sources following the example of Colbert et al. (1996a) (n.b. five sources overlap between their sample of edge-on Seyferts and the present survey). Because of angular resolution limitations, these estimates are accurate only to order of magnitude (plus the implicit uncertainty of the equipartition assumption), and so we present a summary of the results rather than a detailed tabulation for each source. Characteristic equipartition field strengths range over $H_{eq}$ ~ 2 – 20 µG, and total equipartition energies are of order $10^{53} - 10^{54}$ ergs.

Adopting these equipartition values and further assuming energy losses by synchrotron radiation and inverse Compton scattering of microwave background photons (after van der Laan & Perola 1969), the maximum lifetimes of the synchrotron particles (since their last acceleration) range from 3 – 30 Myr. Removing the equipartition assumption, the loss timescale is maximized at field strength $H = H_R / 3^{1/2}$, where $H_R$ is the magnetic field equivalent for the microwave radiation density; $H_R$ ~ 4 µG for the cosmic microwave background at low redshift. Scaling to this maximum loss field strength, the maximum timescale is therefore $\tau_{max} = 29 \, (\nu / 5 \, \text{GHz})^{-1/2}$ Myr. Of course, the AGN and any accompanying starburst will produce additional microwave radiation; the microwave luminosity is ~ 10% the bolometric luminosity (Spinoglio et al. 1995; Nenkova et al. 2002 and references therein), and the resulting, equivalent field at a distance of $l_{kpc}$ kiloparsecs from the AGN and (presumed compact) starburst is $H_R$ ~ 8 $(L_{44} \, l_{kpc}^{-1})$ µG, where $L_{44}$ is the bolometric luminosity of the central engine in units $10^{44}$ erg s$^{-1}$. For this equivalent field strength, $\tau_{max}$ drops to ~ 9 Myr.

The minimum outflow speeds, determined by the (projected) length divided by the maximum source lifetime, scales as $V_{flow} > 34 \, (l_{kpc} \, [\nu / 5 \, \text{GHz}]^{1/2})$ km s$^{-1}$ ($V_{flow} > 110$ km s$^{-1}$ for the AGN plus starburst radiation field $H_R = 8$ µG). As was concluded by Colbert et al. (1996a), relativistic outflows are not required to produce the KSR sources.

## 4.2 *The Detection Fraction of KSRs and Confusion with the Radio Disk*

The detection fraction of KSRs using the present observations and data available in the literature is 44% (19 KSRs / 43 targets); as expected, this detection rate is higher than found in other complete surveys performed at higher angular resolution. To illustrate, we summarize in Table 4 the detection fraction of KSRs in the Thean et al. (2000) survey of 12 μm Seyferts, the Kukula et al. survey of the CfA Seyferts (1995), and the Ulvestad & Wilson (1989) distance-limited survey. The statistics for the Kukula et al. and Thean et al. surveys are capped at $cz < 5200$ km s$^{-1}$ for proper comparison with the present survey. The redshift limit of the Ulvestad & Wilson survey is $cz < 4600$ km s$^{-1}$. All of the scales were adjusted to $H_0 = 70$ km s$^{-1}$ Mpc$^{-1}$ for the purpose of computing detection statistics. Of these three surveys, the Ulvestad & Wilson survey resulted in the highest detection fraction owing to the inclusion of observations at lower frequency ($\nu = 1.4$ GHz) with ~ 2″ angular resolution. The other two surveys were performed at $\nu = 8.4$ GHz and ~ 0.25″ resolution, which effectively resolved out the large scale outflows for detection fractions < 7%. This result highlights the importance of including short-baseline (coarse angular resolution) data towards studying Seyfert radio outflows.

Our reported detection fraction is necessarily a lower limit owing to projection effects: KSRs viewed more nearly pole-on will appear foreshortened and may fall below the 1 kpc selection threshold. In addition, the characteristic surface brightness of the KSR radio emission is comparable to the radio emission from the host galaxy disks of the non-KSR Seyferts. This point is illustrated in Figure 5, which compares the distribution of the average, extra-nuclear 5 GHz brightness temperature for the (14 / 16) KSRs and (17 / 19) disk sources that are resolved by the present observations or measurements available in the literature (we assumed a spectral index $\alpha = -0.7$ for KSRs resolved at other frequencies). Only two of the KSRs in the present sample, NGC 1068 and NGC 5929, significantly exceed the range of average brightness temperatures spanned by the disk sources. Therefore, KSRs that project against disk emission would be effectively hidden by confusion. Taking together these limiting effects of projection and confusion, the 44% detection rate is a firm lower limit, and it cannot be ruled out that most, if not all, Seyfert galaxies produce KSRs.

## *4.3  Comparison of KSR Seyferts vs. non-KSR Seyferts*

We examined some basic properties to determine if there are any statistically significant differences that might distinguish the KSR and non-KSR Seyfert galaxies (Table 5) that might lend some clue to their origin. One important caveat is that some KSRs may have been misidentified as non-KSRs owing to orientation effects (Section 4.2), and the statistics may be diluted as a result. Since the present observations have fixed angular resolution, there is additionally potential for redshift bias in the sense that high-redshift KSRs may be under-resolved and low redshift KSRs might be resolved out by missing, short ($u$, $v$) spacings. The resulting bias would therefore limit the redshift range of detected KSR sources relative to the pool of all Seyfert galaxies. To assess the impact of this selection effect, we compared the redshift distributions of the KSR and non-KSR Seyferts. The KS test (Table 5) failed to find a significant difference in the redshift distributions of the two samples, and so there is no statistical evidence that KSRs are artificially found at higher or lower redshifts. There remains nevertheless the selection effect that smaller (< 5 kpc) outflows are missing at higher redshift owing to finite angular resolution, and the KSR detection fraction is necessarily a lower limit.

One explanation for the non-KSR sources might be that their outflows are stunted by interaction with the ISM of the host galaxy (Smith & Norman 1981; Fiedler & Henriksen 1984). To test this possibility, namely, that there might be some characteristic difference between the host galaxies of KSR and non-KSR Seyferts, we compared several relevant host galaxy properties: (1) $L_{12}$ (i.e., the IRAS 12 μm luminosity), which, as described above, acts as a proxy for the bolometric luminosity of the galaxy (Spinoglio & Malkan 1989; Spinoglio et al. 1995); (2) the morphology of the host galaxy; and (3) the luminosity of the CO 1→0 line, which acts as a proxy for the nuclear gas mass (e.g., Helfer et al. 2003 and references therein). CO measurements were available for 18 KSRs and 13 non-KSRs and were collected from Sanders & Mirabel (1985); Heckman et al. (1989); Young et al. (1995); Maiolino et al. (1997); and Papadopoulos et al. (1998). As summarized in Table 5, no significant difference was found for the parent populations of any of these properties. The statistics for the CO measurements should be viewed somewhat cautiously, however, since the published values are more complete for the KSR Seyferts (18 / 19 sources found in the literature) than for the non-KSR Seyferts (13 / 24 sources found)[3]. With the data presently available, we nevertheless conclude that the host galaxy type or the nuclear gas content has no discernable bearing on the occurrence of a KSR in a Seyfert galaxy.

Another possible explanation is that the central engines of KSR Seyfert galaxies may be more luminous and are perhaps able to generate correspondingly powerful outflows. We therefore compared the distributions of (1) the nuclear 10 μm luminosity ($L_{10} = \nu L_\nu$ at 10 μm) and (2) the [OIII] emission line luminosity of the KSR and non-KSR Seyferts as proxies for the central engine luminosity (see, e.g., Maiolino et al. 1995; Xu, Livio, & Baum 1999). The values of $L_{10}$ used for this analysis were small aperture measurements collected from Gorjian et al. (2004) and Maiolino et al. (1995). Measurements of the [OIII] emission line luminosity were gathered mainly from Whittle (1992), Ho, Filippenko, & Sargent (1997a), and Xu et al. (1999), and additional measurements were found in Veron-Cetty & Veron (1986); Rodriguez-Ardila, Pastoriza, & Donzelli (2000); Schmitt et al. (2003); and Bradley, Kaiser, & Baan (2004), each for individual sources.

A KS test found no significant difference in the distribution of [OIII] luminosity (Table 5). In contrast, the average $L_{10}$ is slightly greater for the KSR Seyferts: log $L_{10}$(KSRs) = 44.7, and log $L_{10}$(non-KSRs) = 44.0. A comparison of the luminosity distributions however failed to find a significant difference; formally, the Peto-Prentice statistic assigns a probability $P = 0.06$ that the parent distributions of $L_{10}$ differ for the KSR and non-KSR samples. Most of the galaxies in the sample were selected (in part) by IRAS 12 μm detections, and it is therefore not surprising that the distributions of nuclear 10 μm luminosities agree, provided that the central engine dominates the host galaxy in mid-infrared radiation. In addition, the published 10 μm data are less complete for the present survey than the published [OIII] luminosities, and so the 10 μm statistics may be correspondingly affected. We conclude that the bolometric luminosity of the central engine is not statistically linked to the presence of a KSR.

---

[3] At the time of the preparation of this manuscript, published measurements were generally more complete for the 12μm Seyferts than for the CfA Seyferts. Where the statistics are incomplete for the joint CfA-12μm sample, they are therefore more nearly complete for the 12μm sub-sample.

Our previous studies raised the question of whether Seyfert KSRs might arise from starburst-driven winds instead of AGN-powered jets. IRAS colors can provide a measure of the relative contribution of a starburst or AGN to the bolometric luminosity, with cooler (redder) colors usually signifying dilution from a starburst (Rush et al. 1993; Bicay et al. 1995; Dopita et al. 1998; Hill et al. 2001). Examining the diagrams of Dopita et al. (1998), the color $[100, 12] \equiv \log[S(100\ \mu m) / S(12\ \mu m)]$ most clearly isolates starbursts, and we therefore used this color as a primary diagnostic, accepting the caveat that reddening may mimic starburst dilution. Nevertheless, there is no significant difference in the distribution of $[100, 12]$ for KSR and non-KSR Seyferts; the probability for identical parent populations is $P = 0.26$. We similarly tested other IRAS colors and the "60μm excess" (Spinoglio, Andreani, & Malkan 2002) but again found no significant difference in the color distribution; in the interest of brevity, Table 5 provides the details only for $[100, 12]$. More traditional diagnostics, such as $[60, 25]$ (de Grijp et al. 1985), may not necessarily discriminate starburst dilution since most of the sample is infrared selected rather than optically selected (Condon et al. 1991), and all of the galaxies are already classified Seyferts or LINERs. Genzel et al. (1998) demonstrated that mid—far infrared line ratios provide a better measure of starburst contribution, but such data are not yet available for many of the sources in the present survey.

Star-forming galaxies follow a tight correlation between global far-infrared and radio continuum emission (Sanders & Mirabel 1996), manifest as a small spread of the parameter $q = \log([FIR / 3.75 \times 10^{12}\ Hz] / S_v(1.4\ GHz))$, where $FIR = 1.26 \times 10^{-14}$ $[2.58\ S_v(60\ \mu m) + S_v(100\ \mu m)]$ W m$^{-2}$ (Helou, Soifer, & Rowan-Robinson 1985). The typical value for spiral galaxies is $q = 2.1 \pm 0.1$ (Helou et al. 1985), and $q = 2.35 \pm 0.2$ for luminous infrared galaxies (Sanders & Mirabel 1996). For the purposes of computing $q$ for the present sample, 1.4 GHz radio flux densities were obtained from the NRAO VLA Sky Survey (NVSS) (Condon et al. 1998). We compared the distribution of $q$ for the KSR and non-KSR Seyferts and found that they are not similarly distributed; the (KS) probability for identical distribution is $P = 0.8\%$ (Table 5). The distributions of $q$ are plotted in Figure 6. From inspection, it appears that most of the non-KSR Seyferts fall within the infrared-radio correlation for normal spirals, but the KSR Seyferts show significantly more variation and a tendency for lower values of $q$ (i.e., excess radio continuum relative to far-infrared continuum).

Markarian Seyferts show a similar relative radio excess in comparison with Markarian starbursts (Bicay et al. 1995), which suggests that low $q$ values signify AGN-related processes even for radio-quiet AGNs. Motivated by this result, we examined the distributions of the nuclear 8.4 GHz luminosity ($L_{8.4}$ in units W Hz$^{-1}$) of the AGN (i.e., a high brightness temperature, compact nuclear radio source, if present) based on the VLA A-array measurements of Kukula et al. (1995) and Thean et al. (2001). The nuclear radio sources of KSR Seyferts are more luminous by nearly one order of magnitude: on average, $\log L_{8.4} = 20.96$ for KSR Seyferts and $\log L_{8.4} = 20.00$ for non-KSR Seyferts. The parent distributions are significantly different with Peto-Prentice probability $P =$ 0.9% for a common parent distribution (Table 5). Figure 7 illustrates the distributions of compact radio source luminosity.

In summary, KSR Seyferts are indistinguishable from non-KSR Seyferts based on (proxies for) the central engine luminosities, host galaxy properties (morphology and bolometric luminosity), and nuclear gas mass. They are further indistinguishable in their infrared colors, which, taken with the luminosity measures, suggests no difference in global star-formation rate (or reddening at mid—far infrared wavelengths). KSRs stand out more clearly by their radio properties: unlike non-KSRs, they tend to deviate from the infrared-radio luminosity correlation, and the radio luminosity of nuclear radio source tends to be greater, by an order of magnitude on average, in KSR sources.

### *4.4 The orientation of the radio and optical axes in KSR Seyferts*

Pressure gradients collimate starburst-driven outflows along the symmetry axis of the galaxy disk (e.g., Chevalier & Clegg 1985; Heckman et al. 1990). Therefore, if the KSRs originate from a starburst, the radio outflow axis should preferentially align with the minor axis of the host galaxy (e.g., Colbert et al. 1996a and references therein). Table 6 summarizes the galaxy major axis data for the KSR Seyferts and the angle between the major axis and the KSR axis. The galaxy major axis data were taken from the RC3 (de Vaucouleurs et al. 1991), Schmitt & Kinney (2000), or measurements of the DSS images. Figure 8 shows the distribution of PA differences. Were the PA differences distributed randomly, the expected count would be ~ 2 sources per 10° bin. A slight excess in the 40° – 60° range curiously appears, suggesting that the KSR axis prefers to align at an oblique angle with the galaxy major axis, and a similar result was found by Colbert et al. (1996a). The deviations of any one bin are however consistent with the expectation for a uniform distribution to within Poisson statistics, and the one-sample KS test fails to find any significant deviation from a uniform distribution of PA differences ($P$ = 0.76 that the distribution is uniform). We conclude that the orientation of the KSR axis is effectively random with respect to the optical major axis of the host galaxy (cf. Nagar & Wilson 1999; Kinney et al. 2000), and it seems unlikely that starburst-driven superwinds, at least of the kind observed to occur in starburst nuclei, dominate the Seyfert KSR population.

We also examined the alignment of the KSR axis and the nuclear jet axis. Table 6 includes a list of 100 mas-scale (VLA / MERLIN) and mas-scale (VLBI) measurements of the small-scale radio structure of the KSR Seyferts. Only 12 of the 19 KSR Seyferts are detected and resolved at these scales, and the distribution of the PA differences is plotted in Figure 9. The sign of the PA difference is positive for KSRs that bend away from the small scale axis but toward the galaxy minor axis and negative for bends toward the galaxy major axis. We used the VLA/MERLIN orientation for the small-scale axis over the VLBA orientation where both were available, because we are interested in measuring how well the KSR aligns with the last (known) trajectory of the small-scale jet that might ultimately feed the KSR. Eight of these KSRs appear to align with the small-scale radio jet to better than 30° in projection. The distribution is however consistent with random orientation; formally, according to a one-sample KS test, the probability that the distribution of PA differences is uniform is $P$ = 0.24. It should be noted, however, that, on smaller scales, the 100 pc-scale jets of Seyfert galaxies also orient randomly with respect to their parsec-scale jets (e.g., Middelberg et al. 2004), and so jet misalignments seem actually to be consistent with an AGN origin. We will consider possible explanations for the jet misalignments in the Section 5.2

## 4.5 Seyfert 1 vs. Seyfert 2

Seyfert unifying schemes predict that the outflows in type 1 Seyferts should appear smaller in projection since they are viewed more nearly pole-on (see the discussion in Kukula et al. 1995 and Thean et al. 2001b). To test this prediction, we statistically compared the radio outflow properties of type 1 and type 2 Seyfert nuclei. For the purposes of these tests, intermediate type 1s were included in the type 1 category.

The simplest test is the detection fraction: KSRs were found in 9/20 Seyfert 1s and 10/23 Seyfert 2s. Formally, using the "difference of two proportions" test (Roy et al. 1994 and references therein), there is (not surprisingly, given the nearly equal ratios) no significant difference in the detection fraction. The probability that the detection fractions of the parent populations are equal is 89%. After removing the putative LINERs from the sample, the fractions become 7/17 Seyfert 1s and 10/22 Seyfert 2s; the probability for equal detection fractions becomes 87%, which again indicates no significant difference.

There also appears to be no significant difference in the distribution of maximum projected linear size of the radio sources. To test this hypothesis, we compiled a list of VLA and MERLIN measurements from the present survey for the KSR Seyferts and from the Kukula et al. (1995) and Thean et al. (2000; 2001b) for the non-KSR Seyferts. Six of the 24 non-KSR Seyferts were not detected by the Kukula and Thean surveys, and so we applied a 15″ upper limit based on the present, C-array point source detections. The Peto-Prentice test failed to find a significant difference between the parent populations of the type 1s and type 2s in our sample; the formal probability is 18% that they were drawn from the same parent population and 7% after removing the putative LINERs. These results suggest that there is no significant difference in the true (deprojected) size of the radio outflows, or their orientation relative to the line-of-sight, on kiloparsec scales.

Seyfert 1 KSRs do not show preferential foreshortening by projection, contrary to the prediction of Seyfert unifying schemes. This result can be reconciled if the KSRs have lost memory of the current AGN axis, whether by jet bending or diversion or temporally varying jet orientation. Otherwise, if the KSRs were to follow the Seyfert unifying schemes such that the Seyfert 1 KSRs are viewed more nearly pole-on, this result would imply that Seyfert 1s have much larger deprojected radio outflow sizes to account for the statistical match in projected size. While this possibility cannot be ruled out with the present data, it would require Seyfert 1s to be distinguishable from Seyfert 2s (i.e., they have intrinsically larger radio outflows), contradicting the main tenet of the unifying schemes. The simpler interpretation remains: the geometry of the large scale radio outflows, whether considering (deprojected) linear extent or orientation with respect to the line-of-sight, is indistinguishable between Seyfert 1s and Seyfert 2s.

One concern with the comparison of the projected linear sizes is that the Kukula and Thean surveys were carried out at a higher frequency (ν = 8.4 GHz) using the VLA in its A-configuration. Steep spectrum emission extended over angular scales intermediate between the resolution of those surveys (~ 0.3″) and the present survey (~ 15″) might have been resolved out, and the resulting gap of angular scale may bias the distribution of maximum projected scale. A more ideal test would require matched frequency measurements including observations with the intermediate B-configuration of the VLA. However, insofar as the KSR detection fractions of type 1s and type 2s are identical, each sample should be equally affected by this bias, and the comparison remains fair.

# 5  DISCUSSION

## *5.1  KSRs originate from nuclear jets*

To summarize the main results, KSRs commonly occur in lower luminosity, radio-quiet AGNs (Seyfert and LINER nuclei), and their occurrence is statistically connected to the radio properties of the central engine. Specifically, KSRs tend to arise in Seyferts that (1) deviate from the radio – far infrared correlation, and are more likely to show a global radio excess, and (2) have more luminous radio nuclei. These results argue that KSRs are more commonly fed by jets arising from the central engine. There is no clear statistical connection between the presence of a KSR (or its properties) and measures of nuclear star-formation, although we cannot specifically rule out the possibility that the KSRs of a few sources might arise from a starburst-driven superwind rather than an AGN-powered jet.

The present radiative luminosity of the central engine appears to have no bearing on the presence of a KSR. This result may owe to the dynamical timescale rather than physical disconnection between the central engine and the outflow generator. Based on proper motions studies of a handful of Seyfert galaxies detected by VLBI techniques and considerations of the jet energetics, the (initial) radio outflow speeds of Seyfert radio ejecta are ~ $0.1c$ (Bicknell et al. 1998; Ulvestad et al. 1999; Middelberg et al. 2004). Setting a lower limit to allow for deceleration, it would therefore take $\tau_{dyn}(KSR) > 10^4 - 10^5$ years for the ejecta to travel the length of a KSR. On the other hand, the presence of a KSR is statistically linked to the radio luminosity on 10 pc scales. The radio outflow phase within the central 10 pc must endure $\geq \tau_{dyn}(KSR)$ to ensure a statistical connection between the two scales.

Roy et al. (1998) found that Seyferts with bright, compact radio cores, independent of the presence of any larger scale outflow, also deviate significantly from the far-infrared – radio correlation. They pointed out that the compact radio source alone does not suffice to explain the radio excess but that there must be significant radio emission on scales larger than 0.1″, the detection limit of the interferometer. Our results self-consistently agree: Seyferts with KSRs also deviate from the far infrared-radio correlation and are more likely to have a bright compact radio core than Seyferts without KSRs. Putting these results together, there appears to be a distinct category of Seyferts with both compact radio cores and large-scale radio outflows (i.e., KSRs). One explanation is that the central engine of a "radio Seyfert" is somehow more efficient at generating outflows. On the other hand, the simpler interpretation may be instead that all Seyferts generate large-scale outflows, but they occur only secularly or periodically. In support of this idea, the nucleus of Mrk 348 recently produced a ~ 2 year duration radio flare that resulted in the ejection of a new, pc-scale radio component (Falcke et al. 2000; Peck et al. 2003), similar in behavior to outbursts observed in radio-loud AGNs.

## 5.2 Jet Misalignments

We find that Seyfert jets oddly misalign from sub-kpc to kpc-scale (cf. Kinney et al. 2000; Schmitt et al. 2001, particularly regarding orientation relative to the host galaxy). Two possible explanations are: (1) the intrinsic orientation of the jet generator (i.e., AGN axis) varies, or (2) the jet trajectory is affected by pressure gradients (i.e., jet-ISM interactions) from small to large scales. In the first possibility, the KSR axis would be a fossil of an earlier jet axis. For radio-loud sources, at least, there is good observational evidence that the jet orientation is variable. There are clear cases of jet precession (see Caproni & Abraham 2004; Caproni, Mosquera Cuesta, & Abraham 2004 and references therein), and, viewed in projection, even small-angle precession can give rise to large, apparent jet misalignments. Rapid (i.e., near-Eddington) accretion events may also alter the spin of the central black hole, which may in turn affect the orientation of the radio jet (Rees 1978; Scheuer & Feiler 1996; Natarajan & Pringle 1998); interestingly, Natarajan & Pringle (1998) estimate ~ $10^5$ yrs as the timescale for redirection of the black hole axis, which is comparable to $\tau_{dyn}$(KSR) in the absence of deceleration. Radiative feedback may also alter the orientation of an accretion disk (Pringle 1996; Pringle 1997). Moreover, black hole merger events, resulting perhaps from the consumption of a minor satellite galaxy, might lead to short timescale redirection of the jet axis such as appears to occur in X-shaped radio jets (Merritt & Ekers 2002).

On the other hand, interaction with the circum-nuclear ISM has provided the popular explanation for the stunted and distorted appearance of Seyfert jets. For example, the radio jet of NGC 1068 is diverted by interaction with molecular gas within the central hundred parsecs (Gallimore et al. 1996a; Gallimore et al. 1996b), and, at a distance of ~ 800 pc from the central engine, the jet terminates in a bow shock which plows into the surrounding ISM at > 80 km s$^{-1}$ (Wilson & Ulvestad 1983; Wilson & Ulvestad 1987). Mrk 6 shows similar bow-shock-like structures on kiloparsec scales (Baum et al. 1993; Kukula et al. 1996). The ~ 800 pc jet of NGC 2110, which appears to propagate directly into the disk of the surrounding host galaxy (Gallimore et al. 1999), seems to have been bent into a gentle S-shape by the rotating ISM (Ulvestad & Wilson 1983). Jet-cloud interactions have also been implicated in the severe bending of radio outflows from parsec scales to hundred-parsec scales in other radio-quiet AGNs (e.g., Irwin & Seaquist 1988; Veilleux, Tully, & Bland-Hawthorn 1993; Steffen, Holloway, & Pedlar 1996;

Daigle & Roy 2001; Middelberg et al. 2004). Morphological correspondences between optical emission line filaments within the NLR and radio jet substructure further argues in favor of jet-ISM interaction (e.g., Whittle et al. 1986; Unger et al. 1988; Whittle et al. 1988; Pedlar et al. 1989; Taylor et al. 1989; Taylor, Dyson, & Axon 1992; Bower et al. 1994; Bower et al. 1995; Capetti et al. 1995a; Capetti et al. 1995b; Capetti et al. 1996; Capetti, Axon, & Macchetto 1997; Ferruit et al. 1998; Capetti et al. 1999; Whittle & Wilson 2004).

Which mechanism is primarily responsible for the misalignment of radio axes from parsec to kiloparsec scale? If jets are affected by interactions with the ISM of the host galaxy, then, on average, jets should follow the pressure gradient and bend towards the rotation axis of the host galaxy. Placing this model as it would be viewed in projection, the KSR axis should bend away from the small-scale jet axis and towards the minor axis of the host galaxy. If instead the jet axis itself changes randomly and the jet direction is not significantly affected by the ISM, then the KSR axis should deviate as often away from the minor axis as toward.

To test this simple model, we evaluated whether the apparent, projected jet bending angles were toward or away from the galaxy minor axis. For the 12 KSR Seyferts with resolved, smaller scale radio structure (see Table 6 & Figure 9), only half showed large scale radio structure bending towards the minor axis of the host galaxy. These results do not rule out temporal, random variations of the intrinsic AGN axis, but they also do not exclusively support that model owing to various diluting effects. For example, projection effects weaken the test because the minor axis does not necessarily correspond to the rotation axis in nearly face-on galaxies (such as Mrk 348). On that note, the four KSRs that show the strongest bending angle toward the minor axis are also the most highly inclined among the KSRs (NGC 3079, NGC 5506, NGC 2992, & NGC 4388) and are less susceptible to this projection effect. In addition, single or few cloud interactions (e.g., NGC 1068: Gallimore et al. 1996a; Gallimore et al. 1996c) may well result in random bending angles. Unfortunately, given the small sample and the fact that there is no strong tendency to bend toward the minor axis except among the more edge-on sources, it is impossible to distinguish the two misalignment mechanisms based solely on this bending angle argument.

## *5.3 Why are Seyfert jets so compact?*

The detection statistics constrain the duty cycle of the outflow phase, $\eta$ (= duration of outflow / average time between outflows) $\geq 44\%$. The characteristic lifetime for Seyfert activity is $\tau = \tau_8 \times 10^8$ yrs (Woltjer 1959; Sanders 1984; Fabian 1999; Mouri & Taniguchi 2002). Assuming *n* outflows occur over this lifetime, the typical duration of any single outflow is therefore $\tau_{\text{flow}} = 44\ (\tau_8\ n^{-1})$ Myr. Irrespective of the site of the most recent particle acceleration, there should be detectable synchrotron emission across a substantial fraction of the KSR, because the energy loss timescale for synchrotron particles compares to the maximum duration of the outflow phase: $\tau_{\text{loss}} \sim 10 - 30$ Myr for electrons maximally radiating at 5 GHz, assuming $B = 10$ μG (i.e., near equipartition for typical brightness temperatures; see Section 4.1).

Given this age constraint, the KSRs must advance with average speed $V_{flow} = 22$ ($l_{kpc}$ $n$ $\tau_8^{-1}$) km s$^{-1}$, where $l_{kpc}$ is the maximum extent of the KSR with respect to the central engine in units kpc ($l_{kpc}$ ranges from 1 to 12 in the present sample). This outflow speed agrees reasonably with the minimum value based on energy loss timescales (Section 4.1). It is however much less than the jet speeds inferred by proper motions, $V_{jet} \sim 0.1c$ and the shock-driven motions of NLR clouds, $V_{NLR} \sim 500 - 1000$ km s$^{-1}$ (Bicknell et al. 1998; Capetti et al. 1999; Wilson & Raymond 1999).

Updating the arguments of Sanders (1984), there are three possible explanations: (1) the radio ejecta disperse after breaking out of a confining medium or magnetic field, and the synchrotron emissivity falls below detectability on larger scales (i.e., the observed size, expressed as $l_{kpc}$, is a lower limit to the true outflow size); (2) the ejecta decelerate by a factor of ~ 1000 over the length of the KSR, presumably by interaction with the NLR and subsequent entrainment of thermal gas or disruption by Rayleigh-Taylor instabilites (e.g., De Young 1986; Bicknell et al. 1998; Brüggen & Kaiser 2001); or (3) the central engine produces many ($n \sim \tau / \tau_{dyn}$[KSR] ~ 1400 $\tau_8$ $l_{kpc}^{-1}$) outflow events over its lifetime. Argument (1) can be essentially ruled out because the KSRs in well-resolved sources terminate in confined lobes or edge-brightened loops (see the discussion of individual sources in Section 3.1). Arguments (2) and (3) otherwise resemble the "frustration" vs. "youth" models, respectively, for gigahertz peaked-spectrum and compact steep-spectrum sources (see O'Dea 1998 for a review). The key difference for the "youth" model is that the relatively high detection fraction of KSRs among Seyfert galaxies requires many outflows over the lifetime of the source; for this reason, we refer to argument (3) instead as the "roman candle" model.

The energy loss timescale poses a serious difficulty for the roman candle model. Assuming no deceleration along the length of the outflow (i.e., a pure roman candle model unaffected by ram pressure or entrainment) and outflow speed = $0.1c$ (to match the observed, pc-scale proper motions), the period between outflows must be only $\tau_{flow}$ / $n$ = 0.07 ($l_{kpc}$) Myr to match the detection statistics. Since the energy loss timescale for synchrotron particles is $\tau_{loss}$ ~ 9 Myr (Section 4.1), the plasma from ~ 130 outflow events should still be radiating significantly at GHz frequencies, and the apparent size of a KSR (i.e., the size of the largest visible structure) would be ~ 550 kpc, assuming two-sided ejection, and there should be dozens of nested radio structures; in contrast, Mrk 6, arguably the most complicated Seyfert radio source, shows only three nested structures within 15 kpc. Reconciling the roman candle model would require increasing the synchrotron loss rate by increasing the characteristic magnetic field to $H_{jet}$ ~ 320 μG, or about a factor of up to 100 above the equipartition field strength. Alternatively increasing the Compton loss rate would require microwave energy densities of order $\omega_R$ ~ $10^{-10}$ ergs cm$^{-3}$, whereas the observed energy densities are two orders of magnitude less: $\omega_R$ ~ $10^{-12}$ $L_{44}$ $l_{kpc}^{-2}$.

p. 23

Although the roman candle model may yet apply to individual sources (requiring detailed modeling beyond the scope of the present work), it seems unlikely to be the typical case. It seems instead more likely that the radio plasma in KSRs must be significantly decelerated along its path, which all but requires interaction with the nuclear ISM (e.g., Wilson & Willis 1980; Taylor et al. 1989). We note that the KSR lobes should travel at least as fast as the expected terminal speed of a buoyant bubble rising out of a galaxy,

$$V_{\text{buoyant}} \sim V_{\text{rot}} \sqrt{\left(\frac{r}{l}\right)\left(\frac{8}{3C}\right)} \qquad (1)$$

where $r$ is the radius of the bubble (i.e., KSR lobe), $l$ is the distance from the center of the gravitational potential, $V_{\text{rot}}$ is the local, Keplerian orbital speed, and $C$ is the drag coefficient (of order unity, but depends on the Reynolds number and the shape of the bubble) (Gull & Northover 1973; Brüggen & Kaiser 2001). Scaling to properties appropriate for KSR lobes ($r \sim 50$ pc, $l \sim 1$ kpc, $v_{\text{rot}} \sim 200$ km s$^{-1}$) gives $V_{\text{buoyant}} \sim 70$ km s$^{-1}$. The inferred advance speeds of the KSR lobes (tens of km s$^{-1}$) compare favorably to this minimum, buoyant speed, suggesting that the lobes receive negligible acceleration by incoming jet plasma.

## 5.4 *Implications of the frustrated jet model*

Given that the jets must decelerate by a large factor, i.e., from $0.1c$ down to lobe advance speeds of tens of km s$^{-1}$, it is of consequent interest to explore the potential impact on the nuclear ISM. For the purposes of the following discussion, we assume that a typical Seyfert only goes through a single radio outflow phase ($n = 1$).

The usual assumption of ram pressure balance leads to a constraint on the plasma density of the jet that feeds the KSR lobes: $\rho_j > 5.4 \times 10^{-31}$ ([$\rho_{\text{ISM}}$ / $10^{-24}$ g cm$^{-3}$] $l_{\text{kpc}}^2$ $\tau_8^{-2}$ [$V_{\text{jet}}$ / $0.1c$]$^{-2}$) g cm$^{-3}$. This estimate is a lower limit, since the jet has probably decelerated below its initial (VLBI-scale) speed before entering a lobe. The kinetic luminosity of the jet is $L_j \sim \pi r_j^2 \rho_j V_{\text{jet}}^3$, or $< 10^{39}$ ([$\rho_{\text{ISM}}$ / $10^{-24}$ g cm$^{-3}$] $l_{\text{kpc}}^2$ [$r_j$ / 50 pc]$^2$ $\tau_8^{-2}$ [$V_{\text{jet}}$ / $0.1c$]) ergs s$^{-1}$ (consistent with the more sophisticated formulation of Bicknell, Dopita, & O'Dea 1997). Here, the radius of the jet, $r_j$, has been limited by the scale of the KSR lobes, $r_j \sim 50$ pc. On 10—100 pc scales, the kinetic luminosity of Seyfert jets is variously estimated as $L_j \sim 10^{42} - 10^{43}$ erg s$^{-1}$ (e.g., Wilson 1997; Bicknell et al. 1998; Capetti et al. 1999). Therefore, the jet must transfer virtually all of its power to the nuclear ISM within the inner kiloparsec, and the residual power of the KSR is relatively feeble, in line with the terminal velocity argument in Section 5.3, above. This result is further consistent with the result that the jet may affect the motions of clouds within the NLR (e.g., Whittle 1985; Taylor et al. 1992; Nelson & Whittle 1996; Steffen et al. 1997; Wilson & Raymond 1999; Dopita et al. 2003) and perhaps contributes significantly to the ionization of NLR clouds (Bicknell et al. 1998; Dopita 2002), although the latter notion is under debate (e.g., Wilson & Raymond 1999; Young, Wilson, & Shopbell 2001).

Based on the present analysis, small scale jets can potentially dump ~ $10^{57}$ ergs into the NLR over their lifetime (cf. Capetti et al. 1999). For comparison, the total energy carried by a starburst-driven superwind is ~ $10^{53}$ – $10^{57}$ ergs (Veilleux 2004); it appears that Seyfert radio jets feed as much or more mechanical energy back to the nuclear ISM as does a nuclear starburst. The amount of material affected by such heating can be estimated based on entrainment arguments. The head of a KSR lobe will directly entrain ISM at a rate $\dot{M}(\text{entrained}) \sim \rho_{ISM} V_{flow} \pi R_{lobe}^2$. Adopting characteristic values $R_{lobe}$ = 100 pc, $\rho_{ISM}$ = $10^{-24}$ g cm$^{-3}$ (Bosma 1981; Cox & Reynolds 1987; Gallimore et al. 1999), and $V_{flow}$ = 22 km s$^{-1}$, we find $\dot{M}(\text{entrained}) \sim$ 0.01 M$_\odot$ yr$^{-1}$. Integrated over the lifetime of the KSR source (and reintroducing normalizations), the total mass swept up by a KSR is $M(\text{entrained}) \sim 4.6 \times 10^5$ ([$\rho_{ISM}$ / $10^{-24}$ g cm$^{-3}$][$R_{lobe}$ / 100 pc]$^2$ $l_{kpc}$) M$_\odot$. Steady-state transfer of momentum will increase the total mass of entrained material by a factor of ~ 2 (De Young 1986). For comparison, the mass of the atomic ISM is of order $10^7$ M$_\odot$ within the inner kiloparsec of Seyfert galaxies (Gallimore et al. 1999). Therefore, although Seyfert radio outflows provide a significant source of energy to the nucleus of a spiral galaxy, they affect only ~ 10% of the total nuclear ISM.

The key result therefore is that the jet loses the bulk of its power in the inner kiloparsec, and, because the initial jet motion is highly supersonic in the surrounding ISM, energy transfer to the ISM occurs mainly by fast shocks. Shock-heating of the ISM results in immediate post-shock temperatures of order $10^6$ – $10^7$ K (Taylor et al. 1992), and so the primary jet shocks should therefore produce copious soft X-ray emission. For comparison, the luminosity of the extended soft X-ray component of Seyfert galaxies is of order $10^{41}$ – $10^{42}$ erg s$^{-1}$ (e.g., Weaver et al. 1995 and references therein; Young et al. 2001), or roughly 10% of the energy that must be lost by the radio jet within the NLR. Shock models indeed predict that ~ 10% of the kinetic power of jet-induced shocks is radiated in soft X-rays (Dopita & Sutherland 1996; Wilson & Raymond 1999). We conclude that the "frustrated jet" model explains well the loss of jet luminosity from sub-kpc to kpc scales as well as the compactness of Seyfert jets.

# 6 CONCLUSIONS

The primary results of this survey follow.

1. Kiloparsec-scale radio outflows (KSRs) in nearby AGNs (Seyfert and LINER galaxies) are common, with a detection rate of 44% for the present, distance-limited subset of the combined 12μm and CfA Seyfert samples.
2. The brightness temperature of the radio emission from the kiloparsec scale lobes compares with the emission from the star-forming disks, and so the detection of kiloparsec-scale outflows is limited both by projection and confusion. The detection rate of 44% is therefore a lower limit.
3. Seyfert KSRs are randomly oriented with respect to the galaxy minor axis and the sub-kpc radio jet axis.

4. Seyferts with KSRs deviate from the far infrared-radio correlation, tending toward radio excess, and they tend to have more luminous compact nuclear radio sources than Seyferts lacking KSRs. The global infrared properties of Seyferts with and without KSRs are moreover indistinguishable, indicating that the presence of a KSR is not (statistically) linked to the presence of a starburst. Consistent with the findings of Colbert et al. (1996a), KSRs are more likely generated by an AGN-powered jet than a starburst-driven superwind.
5. The detection rate of KSRs implies an outflow time-scale $\tau_{flow}$ ~ 44 Myr, but the dynamical time is only $\tau_{dyn}$ ~ 0.1 Myr assuming jet speeds inferred from proper motion studies. The likeliest explanation is that the radio outflows are strongly decelerated by interaction with the nuclear ISM (the "frustrated jet" model).
6. The time-scale issue might also be resolved if the outflows may occur in short bursts (the "roman candle" model). However, the required period between outflow bursts (~ 70 kyr) is much shorter than the energy loss timescale for synchrotron particles (~ 9 Myr, allowing for strong inverse Compton cooling), which in turn is still much longer than $\tau_{dyn}$. Therefore, the roman candle model still requires strong deceleration to account for the apparent compactness of Seyfert jets.
7. Based on ram pressure and terminal velocity considerations, the kinetic power of the jet feeding the lobes of a KSR must be orders of magnitude less than the jet power on 10—100 pc scales. Therefore, virtually all of the jet power of a Seyfert nucleus is lost to the nuclear ISM within the inner kpc. The luminosity of extended soft X-ray emission in Seyfert nuclei is observed to be ~ 10% of the jet energy lost, consistent with the predictions of shock models. This result supports the "frustrated jet" model.
8. Seyfert radio sources may provide ~ $10^{57}$ ergs to the nuclear ISM through shocks over their lifetimes, comparable to that provided by starburst-driven superwinds. The resulting KSRs may entrain (or otherwise stir up) ~ 10% of the neutral, atomic ISM within the inner kpc.

Although the present results favor deceleration (or, frustration) by jet-ISM interactions to explain the too short dynamical ages of the Seyfert radio sources, spectral aging measurements derived from lower frequency observations may further test this scenario against the roman candle model. Since the frustration model predicts that the KSR lobes are traveling near the slow, terminal speed of a buoyant lobe with little energy input from the jet, we would expect little rejuvenation of the synchrotron electron energies. Given the large detection fraction of KSR sources, the frustration model therefore predicts strong spectral steepening (e.g., Myers & Spangler 1985) corresponding to a synchrotron age comparable to the age of the AGN (of order $10^8$ years). The roman candle model instead predicts relatively young synchrotron plasma.

**Acknowledgements**

We acknowledge helpful conversations with Martin Elvis and Moshe Elitzur. We also thank the anonymous referee for suggestions that helped to clarify the discussion of the sample selection and the following statistical analyses. This research has made use of the

# Figures

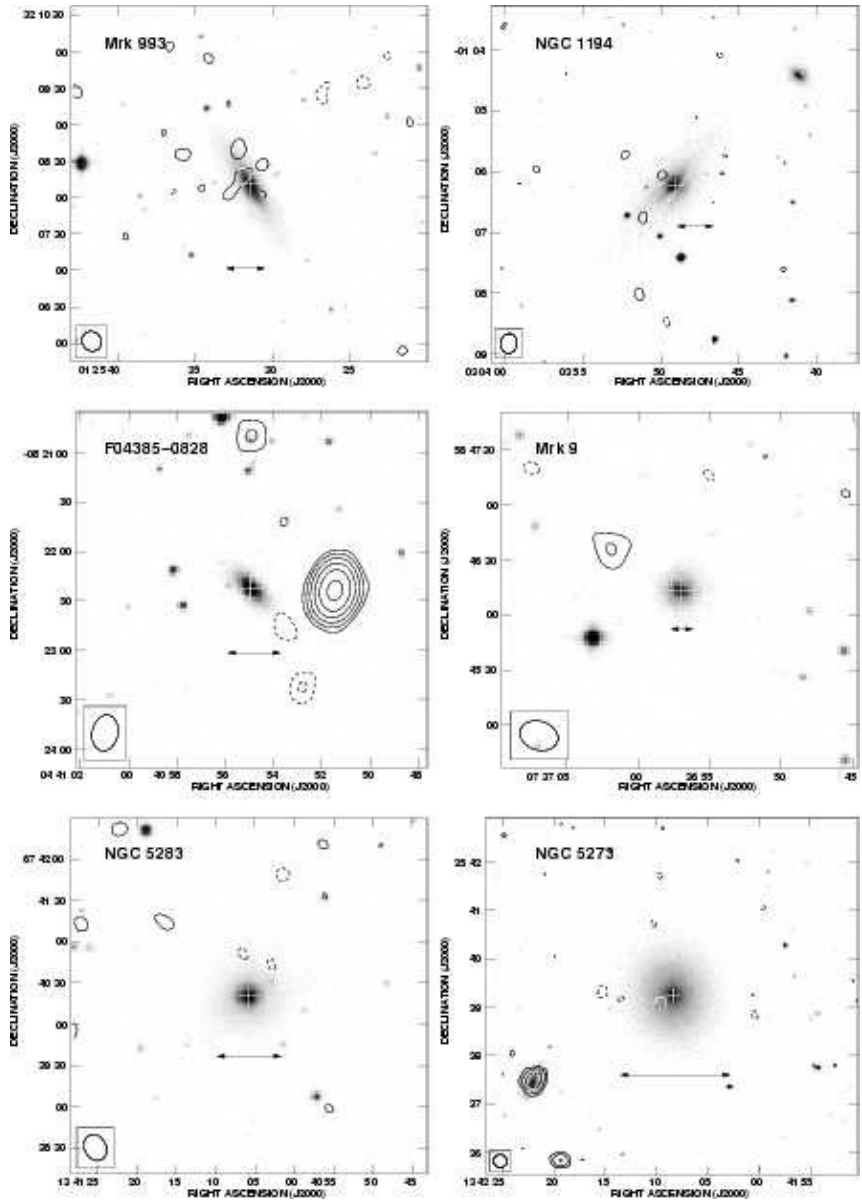

**Figure 1:** Seyfert galaxies classified as point sources (at the 15″ – 20″ resolution of the present survey). The 5 GHz radio continuum is plotted as contours, and the contour levels are $3\sigma \times (-2, -1, 1, 2, 4, 8, \ldots, 512)$, where $\sigma$ is the image RMS listed in Table 1. A point source has been subtracted at the position of the "+" symbol. The horizontal, dual-direction arrows are 10 kpc scale bars. The Seyfert identification is annotated in the upper left corner of each panel.

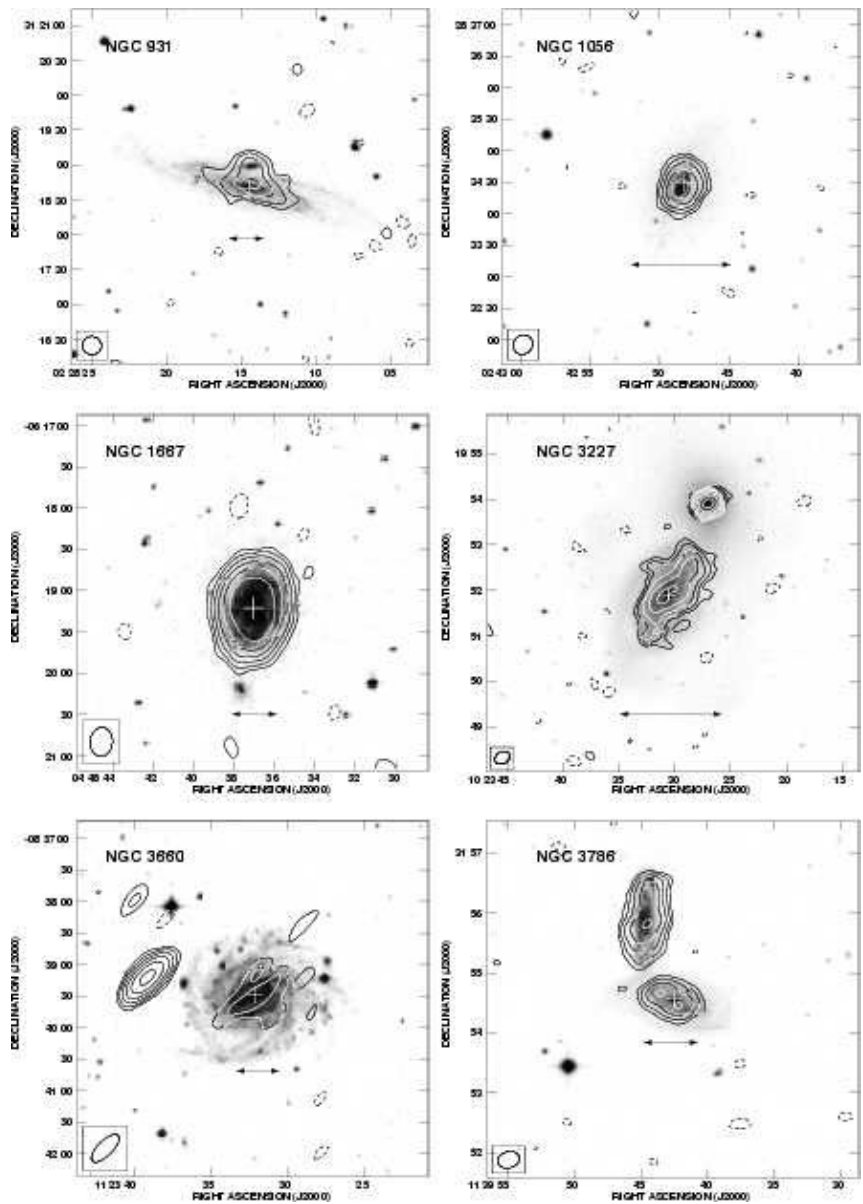

**Figure 2: Seyfert galaxies classified as galaxy disk sources. The plotting convention is identical to that in Figure 1.**

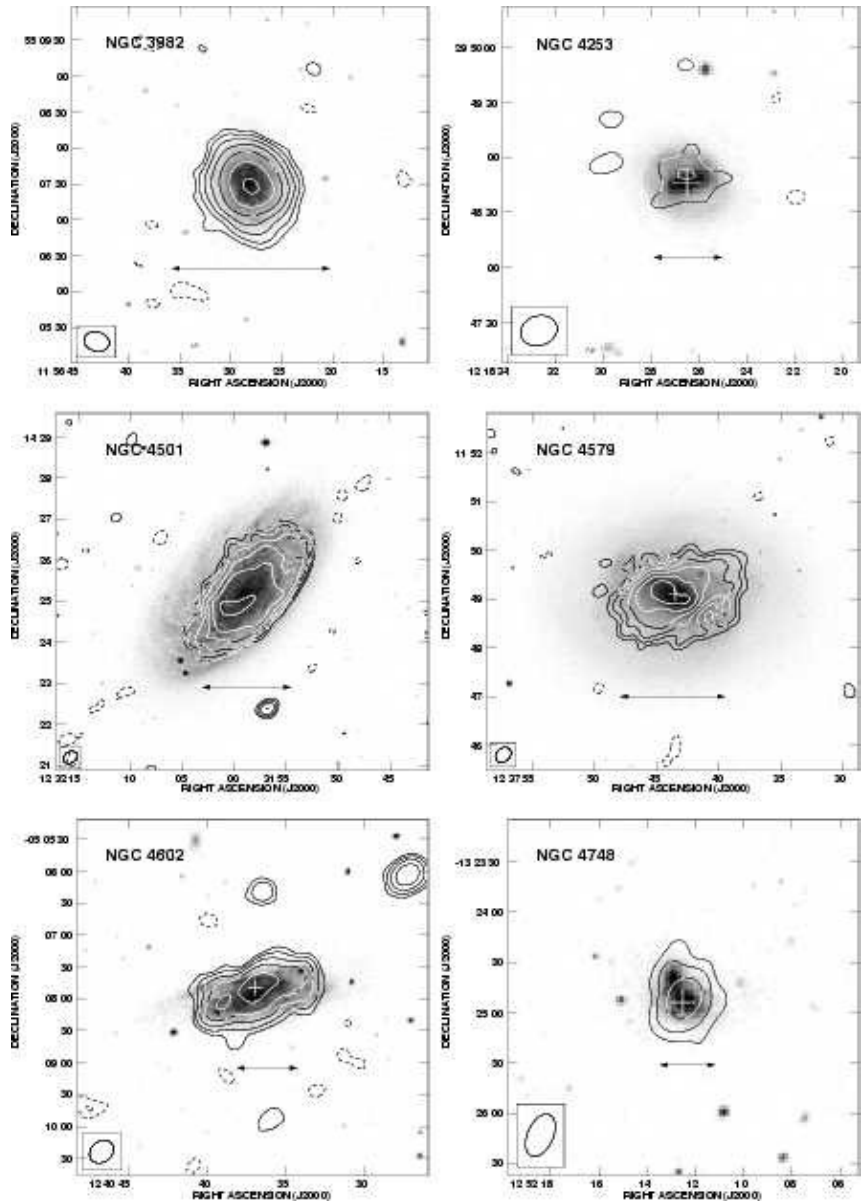

(Figure 2, continued)

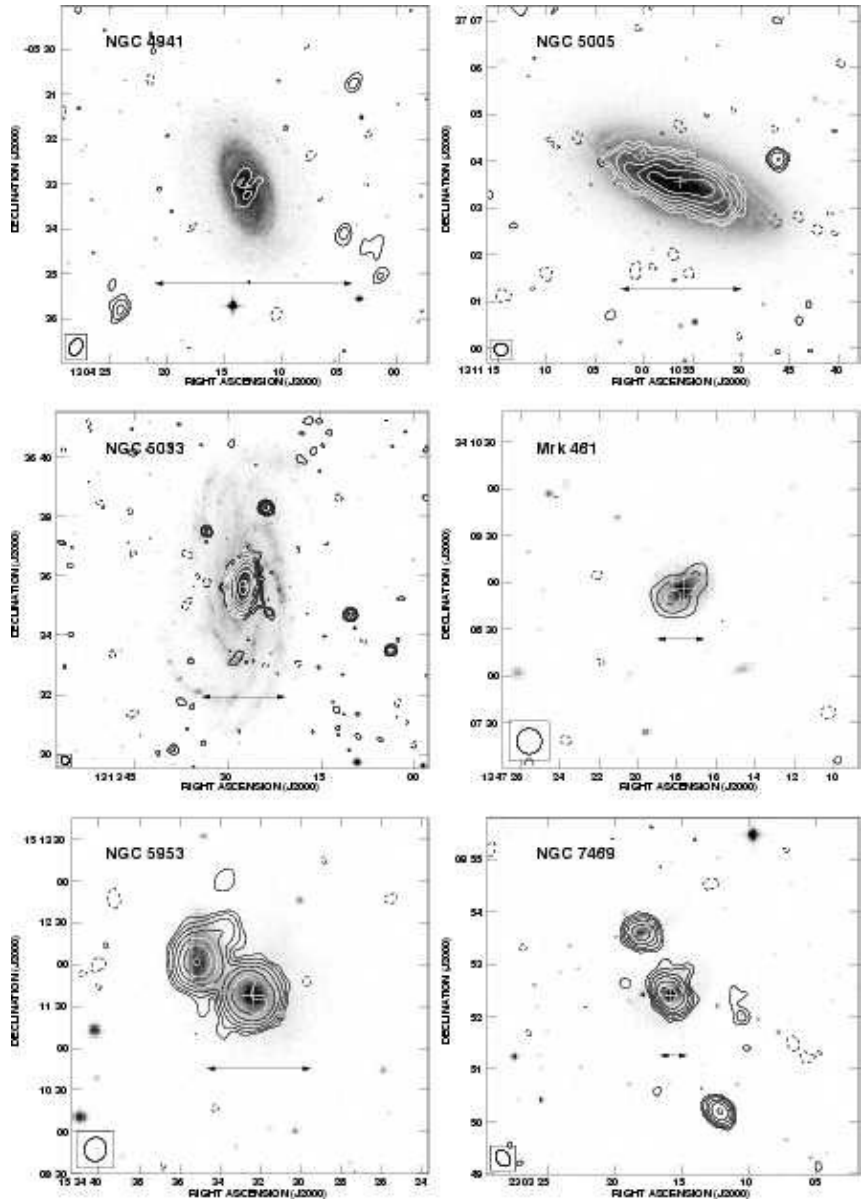

(Figure 2, continued)

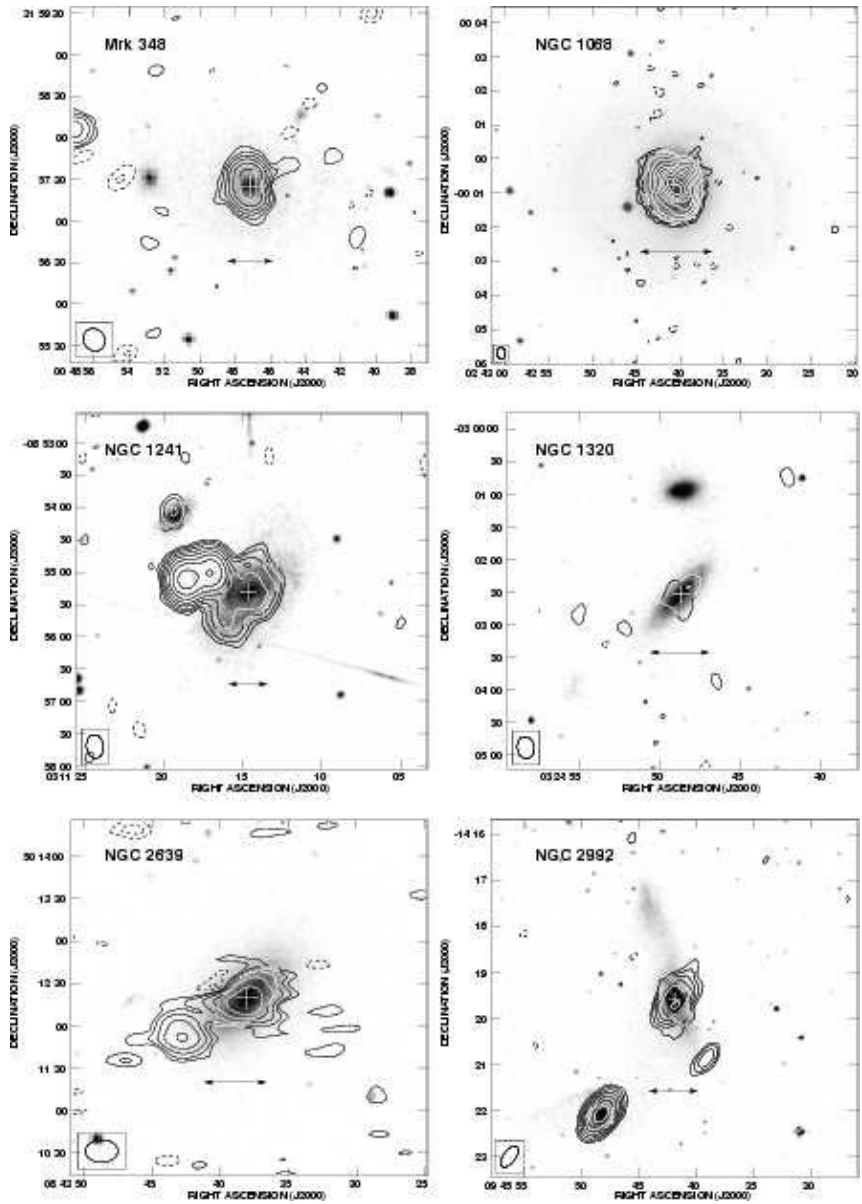

**Figure 3:** Seyfert galaxies classified as candidate KSR sources based either on the present observations or observations in the literature. The plotting convention is identical to that in Figure 1. Note that two point sources were subtracted from the radio image of NGC 5929 to reveal the underlying disk emission. KSR sources not resolved by the present observations include NGC 1241, NGC 2992, NGC 5506, and NGC 5929.

p. 38

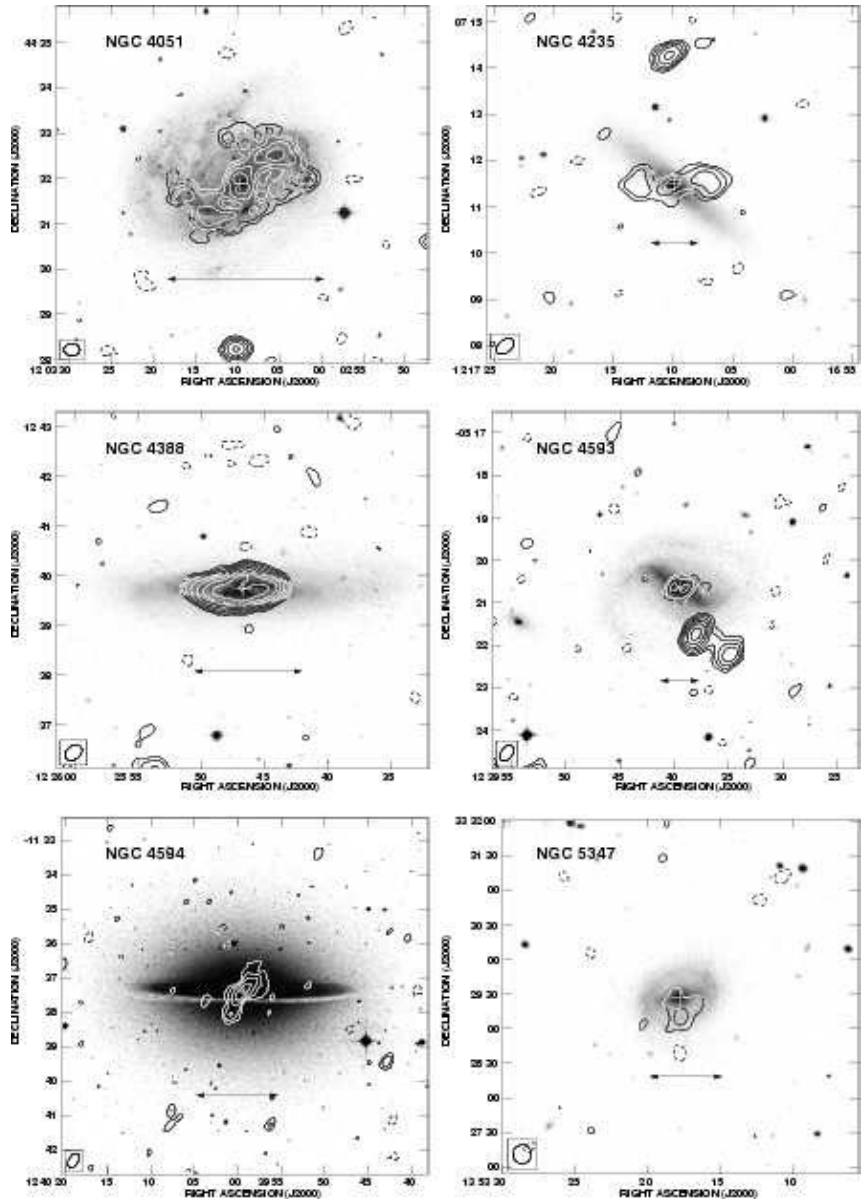

(Figure 3, continued)

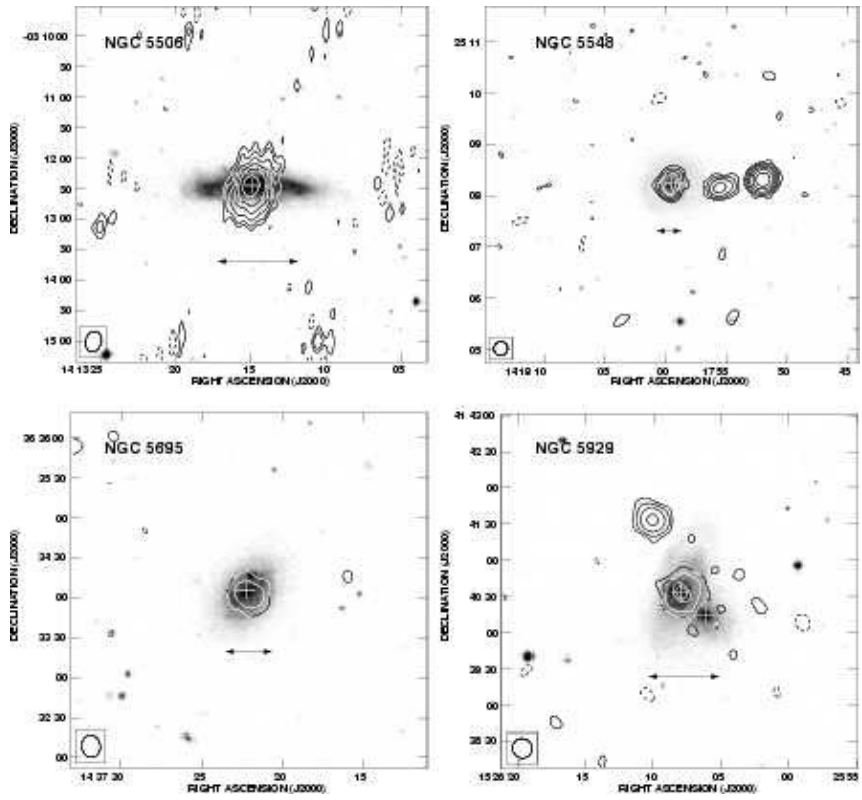

(Figure 3, continued)

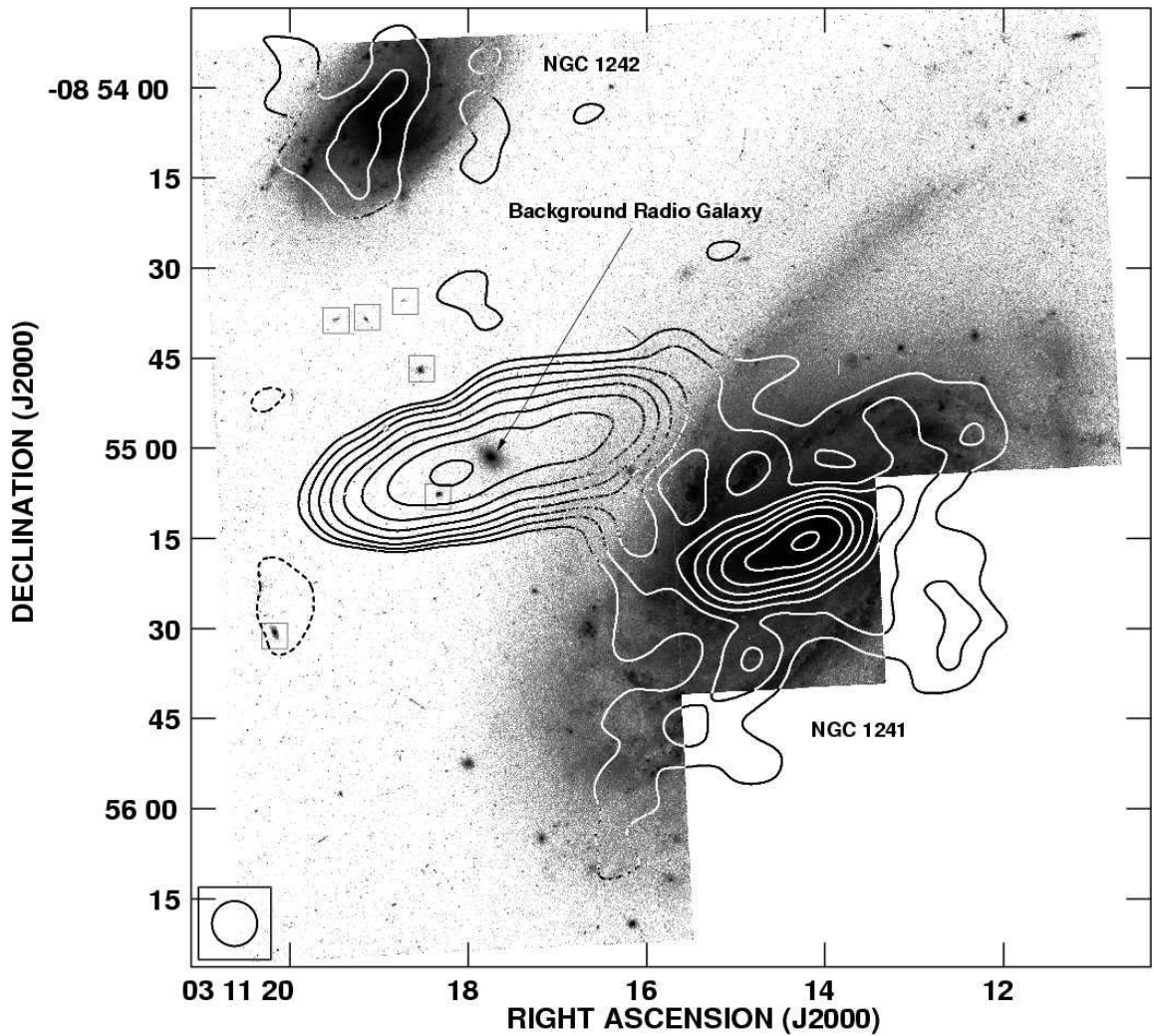

**Figure 4: A detailed view of the Seyfert galaxy NGC 1241. An archival HST/WFPC-2 image, taken with the F606W (~ 600 nm) filter, is shown as grayscale with a histogram stretch to emphasize the faint background sources. The archival 1.5 GHz VLA image is shown as overlaid contours. The contour levels are ±0.24, 0.39, 0.63, 1.0, 1.7, 2.7, 4.5, and 7.3 mJy beam$^{-1}$. The open circle in the lower left corner traces the 7.5″ CLEAN restoring beam. Possible background galaxies are marked by open squares, and the candidate background radio galaxy is so annotated.**

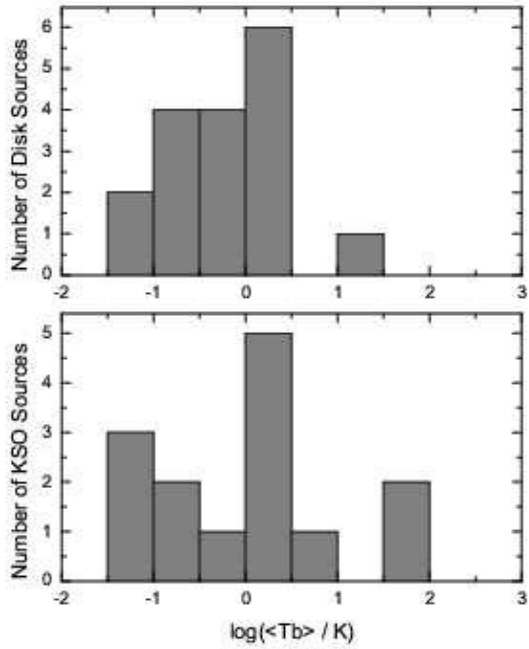

**Figure 5:** Distribution of $<T_b>$, the average 5 GHz brightness temperature of disk emission in resolved disk sources (top panel) and outflow emission in resolved KSR sources. Since the surface brightness of radio emission from disks compares to the surface brightness of KSRs, KSRs may have been missed in disk sources owing to confusion.

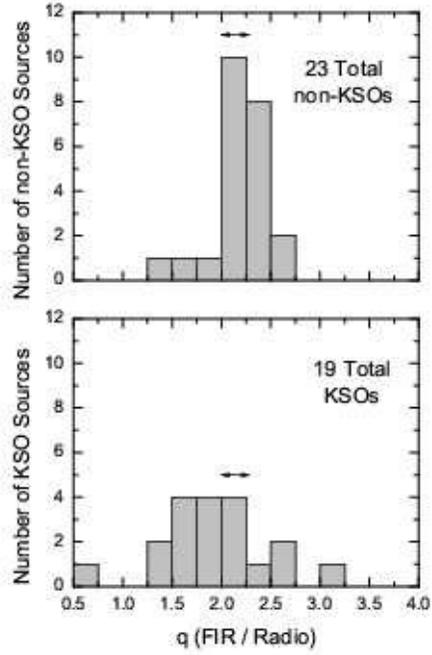

**Figure 6. A comparison of the distribution of *q* (far-infrared / radio continuum ratio) for non-KSR Seyferts (top panel) and KSR Seyferts (bottom panel). The narrow range of *q* characteristic of normal spiral galaxies is indicated in both panels by the horizontal double arrows.**

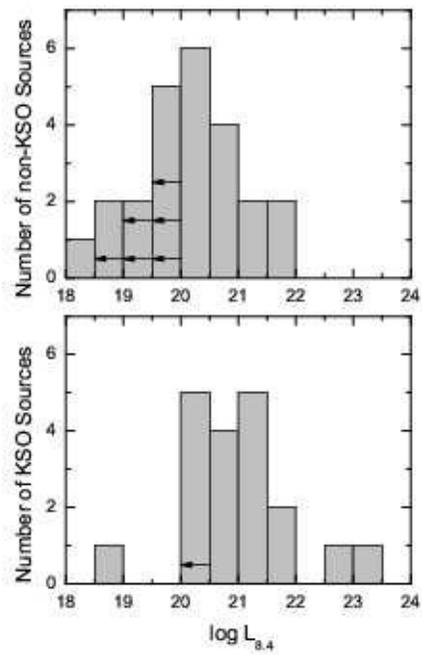

**Figure 7 .** A comparison of the distribution of the luminosity of the nuclear radio source (8.4 GHz luminosity in W Hz$^{-1}$) for non-KSR Seyferts (top panel) and KSR Seyferts (bottom panel). Leftward pointing arrows mark bins containing upper limits.

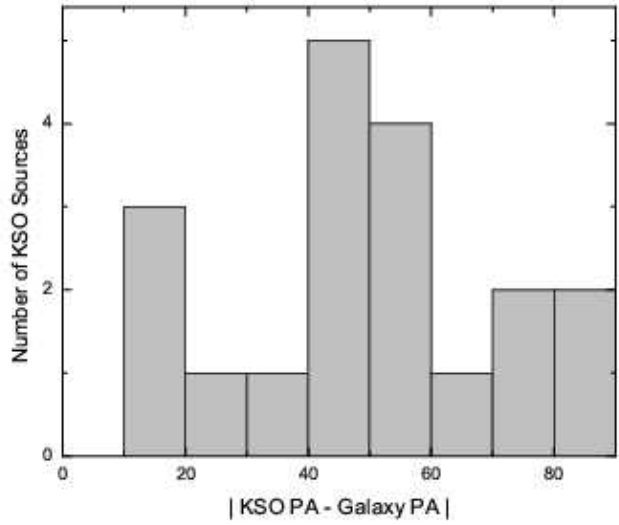

**Figure 8:** Histogram of the position angle difference in degrees between the KSR axis and the host galaxy major axis for the 19 KSR Seyferts found in the present sample.

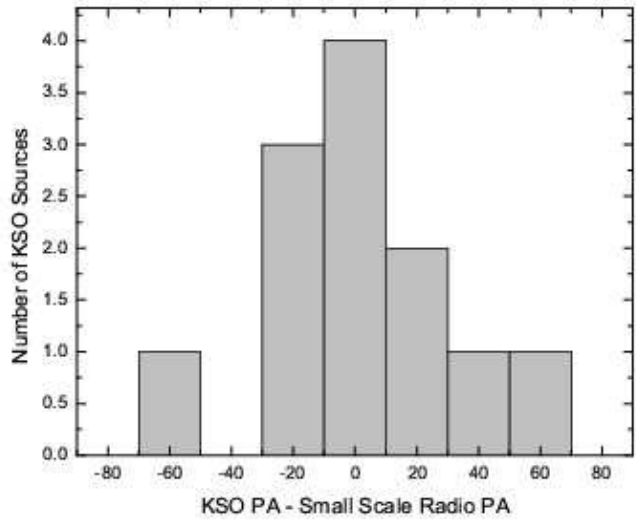

**Figure 9: Histogram of the position angle difference in degrees between the KSR axis and the small scale radio axis for the 12 out of 19 KSR Seyferts that have resolved, small scale radio structure. The sign convention is positive for KSRs that tend to bend toward the galaxy minor axis and negative for bends toward the galaxy major axis.**

# Tables

**Table 1 Observed Seyfert Galaxies**

| Source | Other ID | Type | Hubble Type | $cz$ | RMS ($\mu$Jy beam$^{-1}$) | Beam Size (FWHM) ($'' \times ''$, °) | | |
|---|---|---|---|---|---|---|---|---|
| Mrk 348 | NGC 262 | 2 | 0 | 4507 | 39.4 | 16.8 | 15.0 | 39 |
| Mrk 993 | | 2 | 1 | 4658 | 36.6 | 16.7 | 15.3 | 42 |
| NGC 931 | Mrk 1040 | 1 | 4 | 4992 | 36.9 | 16.3 | 15.7 | 72 |
| NGC 1068 | M 77 | 2 | 3 | 1137 | 97.6 | 19.8 | 15.7 | 11 |
| NGC 1056 | Mrk 1183 | 2 | 1 | 1545 | 75.2 | 19.4 | 17.5 | −40 |
| NGC 1194 | | 1 | −0.5 | 4057 | 83.8 | 19.5 | 15.5 | −6 |
| NGC 1241 | | 2 | 3 | 4052 | 42.7 | 21.8 | 15.9 | 7 |
| NGC 1320 | Mrk 607 | 2 | 1 | 2716 | 42.7 | 20.2 | 15.9 | 11 |
| F04385-0828 | | 2 | −2 | 4527 | 41.8 | 21.9 | 15.9 | −12 |
| NGC 1667 | | 2 | 5 | 4547 | 39.7 | 21.3 | 16.3 | −5 |
| Mrk 9 | | 1.5 | −2 | 11954 | 62.7 | 21.1 | 15.7 | 72 |
| NGC 2639 | | 1.9 LINER | 1 | 3336 | 41.5 | 22.8 | 15.9 | −89 |
| NGC 2992 | | 2 | 1 | 2311 | 82.0 | 31.6 | 15.3 | −35 |
| NGC 3227 | | 1.5 | 1 | 1157 | 41.1 | 20.4 | 15.8 | −57 |
| NGC 3660 | | 2 | 4 | 3678 | 63.5 | 34.7 | 14.1 | −46 |
| NGC 3786 | Mrk 744 | 1.8 | 1 | 2678 | 40.0 | 21.4 | 16.1 | −73 |
| NGC 3982 | | 2 | 3 | 1109 | 42.8 | 21.5 | 16.0 | 75 |
| NGC 4051 | | 1.5 | 4 | 700 | 41.7 | 20.4 | 15.9 | 89 |
| NGC 4235 | | 1 | 2 | 2410 | 43.7 | 25.8 | 16.0 | −51 |
| NGC 4253 | Mrk 766 | 1.5 | 1 | 3876 | 55.1 | 20.1 | 16.1 | −71 |
| NGC 4388 | | 2 | 3 | 2524 | 34.3 | 21.8 | 16.0 | −51 |
| NGC 4501 | M 88 | 2 | 3 | 2281 | 45.4 | 20.8 | 15.9 | −51 |
| NGC 4579 | M 58 | 1.9 LINER | 3 | 1519 | 42.9 | 20.6 | 15.9 | −47 |

| Source | Other ID | Type | Hubble Type | $cz$ | RMS ($\mu$Jy beam$^{-1}$) | Beam Size (FWHM) ($'' \times ''$, °) | | |
|---|---|---|---|---|---|---|---|---|
| николаев NGC 4593 | Mrk 1330 | 1 | 3 | 2698 | 39.8 | 24.9 | 15.7 | −33 |
| NGC 4594 | M 104 | 1.9 LINER | 1 | 1024 | 50.3 | 26.9 | 15.7 | −29 |
| NGC 4602 | | 1.9 | 4 | 2539 | 50.6 | 24.6 | 19.6 | −54 |
| NGC 4748 | | 1 | 1 | 4386 | 50.9 | 26.9 | 15.6 | −25 |
| NGC 4941 | | 2 | 2 | 1108 | 62.9 | 23.1 | 15.7 | −26 |
| NGC 5005 | | 2 LINER | 4 | 946 | 40.8 | 16.4 | 16.0 | 68 |
| NGC 5033 | | 1.9 | 5 | 875 | 40.9 | 16.2 | 16.0 | 43 |
| NGC 5283 | Mrk 270 | 2 | −2 | 3119 | 63.2 | 19.7 | 15.6 | 26 |
| NGC 5273 | | 1.9 | −2 | 1064 | 58.0 | 16.2 | 16.0 | 23 |
| Mrk 461 | | 2 | ≥0 | 4856 | 37.4 | 16.1 | 15.8 | 3 |
| NGC 5347 | | 2 | 2 | 2335 | 36.9 | 16.1 | 15.8 | −1 |
| NGC 5506 | | 1.9 | 1 | 1853 | 39.6 | 20.8 | 15.6 | −13 |
| NGC 5548 | | 1.5 | 0 | 5149 | 68.3 | 16.3 | 15.7 | −12 |
| NGC 5695 | Mrk 686 | 2 | ≥0 | 4225 | 74.1 | 16.7 | 13.6 | 7 |
| NGC 5929 | | 2 | 2 | 2492 | 52.0 | 16.5 | 15.9 | 47 |
| NGC 5953 | | 2 | 1 | 1965 | 38.4 | 17.2 | 15.7 | −17 |
| NGC 7469 | | 1.2 | 1 | 4892 | 47.9 | 20.0 | 14.6 | 35 |

**Table 2 Results of point source subtraction analysis**

| ID | Class | $S_{5\text{ GHz}}$ (point) (mJy) | α(J2000) (h m s) | | | δ(J2000) (° ′ ″) | | |
|---|---|---:|---|---|---|---|---|---|
| Mrk 348 | KSR | 801.7 | 00 | 48 | 47.132 | +31 | 57 | 24.93 |
| Mrk 993 | Point | 2.8 | 01 | 25 | 31.439 | +32 | 08 | 10.62 |
| NGC 931 | Disk | 1.7 | 02 | 28 | 14.545 | +31 | 18 | 40.19 |
| NGC 1068 | KSR | 1342.4 | 02 | 42 | 40.718 | −00 | 00 | 46.80 |
| NGC 1056 | Disk | 4.8 | 02 | 42 | 48.204 | +28 | 34 | 28.83 |
| NGC 1194 | Point | 1.5 | 03 | 03 | 49.071 | −01 | 06 | 14.28 |
| NGC 1241 | KSR | 12.3 | 03 | 11 | 14.658 | −08 | 55 | 18.24 |
| NGC 1320 | KSR | 3.3 | 03 | 24 | 48.698 | −03 | 02 | 32.41 |
| F04385-0828 | Point | 12.7 | 04 | 40 | 54.971 | −08 | 22 | 22.03 |
| NGC 1667 | Disk | 5.0 | 04 | 48 | 37.029 | −06 | 19 | 12.64 |
| Mrk 9 | Point | 1.8 | 07 | 36 | 57.007 | +58 | 46 | 13.36 |
| NGC 2639 | KSR | 119.2 | 08 | 43 | 38.048 | +50 | 12 | 19.91 |
| NGC 2992 | KSR | 91.6 | 09 | 45 | 41.926 | −14 | 19 | 35.20 |
| NGC 3227 | Disk | 35.0 | 10 | 23 | 30.565 | +19 | 51 | 54.07 |
| NGC 3660 | Disk | 0.8 | 11 | 23 | 32.144 | −08 | 39 | 29.03 |
| NGC 3786 | Disk | 7.3 | 11 | 39 | 42.543 | +31 | 54 | 34.01 |
| NGC 3982 | Disk | 17.9 | 11 | 56 | 27.983 | +55 | 07 | 31.00 |
| NGC 4051 | KSR | 9.4 | 12 | 03 | 09.624 | +44 | 31 | 52.54 |
| NGC 4235 | KSR | 7.5 | 12 | 17 | 09.861 | +07 | 11 | 29.35 |
| NGC 4253 | Disk | 20.4 | 12 | 18 | 26.522 | +29 | 48 | 46.65 |
| NGC 4388 | KSR | 34.6 | 12 | 25 | 46.764 | +12 | 39 | 44.77 |
| NGC 4501 | Disk | 2.6 | 12 | 32 | 00.336 | +14 | 24 | 40.52 |
| NGC 4579 | Disk | 40.7 | 12 | 37 | 43.503 | +11 | 49 | 05.86 |
| NGC 4593 | KSR | 3.8 | 12 | 39 | 39.463 | −05 | 20 | 39.03 |
| NGC 4594 | KSR | 2.7 | 12 | 39 | 59.440 | −11 | 37 | 22.99 |
| NGC 4602 | Disk | 1.3 | 12 | 40 | 38.061 | −05 | 07 | 40.51 |
| NGC 4748 | Disk | 6.0 | 12 | 52 | 12.521 | −13 | 24 | 53.82 |
| NGC 4941 | Disk | 9.0 | 13 | 04 | 13.105 | −05 | 33 | 05.46 |
| NGC 5005 | Disk | 15.7 | 13 | 10 | 56.231 | +37 | 03 | 33.23 |
| NGC 5033 | Disk | 6.5 | 13 | 13 | 27.278 | +36 | 35 | 38.84 |
| NGC 5283 | Point | 6.7 | 13 | 41 | 05.871 | +67 | 40 | 20.43 |
| NGC 5273 | Point | 1.9 | 13 | 42 | 08.429 | +35 | 39 | 15.20 |
| Mrk 461 | Disk | 1.9 | 13 | 47 | 17.697 | +34 | 08 | 55.57 |
| NGC 5347 | KSR | 3.1 | 13 | 53 | 17.771 | +33 | 29 | 27.03 |
| NGC 5506 | KSR | 227.2 | 14 | 13 | 14.869 | −03 | 12 | 27.71 |
| NGC 5548 | KSR | 11.2 | 14 | 17 | 59.460 | +25 | 08 | 14.36 |
| NGC 5695 | KSR | 0.8 | 14 | 37 | 22.252 | +36 | 34 | 05.89 |
| NGC 5929 | KSR | 19.6 | 15 | 26 | 07.942 | +41 | 40 | 33.72 |
| NGC 5953 | Disk | 17.4 | 15 | 34 | 32.335 | +15 | 11 | 37.66 |
| NGC 7469 | Disk | 61.6 | 23 | 03 | 15.774 | +08 | 52 | 24.42 |

**Table 3 KSR Properties**

| Source | PA (°) | Extent (″) | Extent (kpc) | $S_5$ (mJy) | Morphology | Sidedness | Ref(s) |
|---|---|---|---|---|---|---|---|
| Mrk 348 | 13 | 18 | 5.6 | 8.5 | Lobes | 2 | 1, 2 |
| NGC 1068 | 33 | 15 | 1.1 | 664.0 | Lobes / Linear | 2 | (3, 4) |
| NGC 1241 | 100 | 16 | 4.5 | 1.2 | Tongue | 1 | 1 |
| NGC 1320 | 186 | 14 | 2.6 | 0.4 | Tongue | 1 | 1, 5 |
| NGC 2639 | 120 | 53 | 12.2 | 2.3 | Lobe / Linear | 1 | 1 |
| NGC 2992 | 102 | 30 | 4.5 | 49.9 | Tongue (smaller lobes) | 1 (2) | 1, 5, 6, 7 |
| NGC 3079 | 63 | 45 | 3.6 | 44.8 | Lobes | 2 | (8, 9, 10) |
| NGC 3516 | 44 | 45 | 8.3 | 7.4 | Lobe / Linear | 1 | (2, 11) |
| NGC 4051 | 3 | 30 | 1.5 | 2.9 | Lobes | 2 | 1, 2 |
| NGC 4151 | 85 | 15 | 1.0 | 128.0 | Linear | 2 | (2, 12) |
| NGC 4235 | 93 | 110 | 10.7 | 3.5 | Lobes | 2 | 1, 5 |
| NGC 4388 | 7 | 40 | 3.1 | 19.0 | Tongues | 2 | 1, 13, 14, 15 |
| NGC 4593 | 91 | 26 | 4.9 | 0.8 | Lobes? | 2 | 1 |
| NGC 4594 | 142 | 54 | 3.8 | 3.1 | Linear | 2 | 1 |
| NGC 5347 | 182 | 16 | 2.6 | 0.5 | Tongue | 1 | 1 |
| NGC 5506 | 140 | 45 | 5.9 | 39.6 | Tongues | 2 | 1, 5, 16 |
| NGC 5548 | −15 | 12 | 4.3 | 3.2 | Lobes | 2 | 1, 2, 17, 18 |
| NGC 5695 | 72 | 18 | 5.3 | 1.3 | Tongue | 1 | 1 |
| NGC 5929 | 172 | 15 | 2.6 | 1.2 | Lobes | 2 | (2) |

1. The present work
2. Baum et al. 1993
3. Wilson & Ulvestad 1983
4. Gallimore et al. 1996c
5. Colbert et al. 1996a
6. Wehrle & Morris 1988
7. Ulvestad & Wilson 1984b
8. Duric et al. 1983
9. Duric & Seaquist 1988
10. Irwin & Saikia 2003
11. Wrobel & Heeschen 1988
12. Pedlar et al. 1993
13. Hummel et al. 1983
14. Condon 1987
15. Hummel & Saikia 1991
16. Wehrle & Morris 1987
17. Wilson & Ulvestad 1982a
18. Wrobel 2000

**Notes** – $S_5$ refers to the flux density of the extranuclear, KSR emission at 5 GHz. "Lobes" morphology refers to radio structure disconnected from the nucleus; "Tongue" refers to a smooth, broad extension of radio emission from the nucleus; and "Linear" refers to jet-like structures. Additional details are given in Section 3.1.

**Table 4 KSR detection statistics from complete surveys of Seyfert galaxies**

| Survey | ν (GHz) | Array | Resolution | $cz_{max}$ (km s$^{-1}$) | N (Total) | N (KSRs) | Detection Fraction |
|---|---|---|---|---|---|---|---|
| Thean et al. (2000) | 8.4 | VLA-A | ~0.25″ | 5200 | 60 | 4 | 6.7% |
| Kukula et al. (1995) | 8.4 | VLA-A | ~0.25″ | 5200 | 18 | 0 | 0 |
| Ulvestad & Wilson (1989) | 1.4 | VLA-A/B | ~ 2″ | 4600 | 57 | 13 | 23% |

**Table 5 KSR vs. non-KSR Comparison Statistics**

| Property | N(KSR) | N(non-KSR) | Test | Value | P |
|---|---|---|---|---|---|
| $cz$ | 19 | 24 | KS | 0.21 | 0.76 |
| Log $L_{12}$ (global) | 19 | 23 | KS | 0.23 | 0.56 |
| Hubble Type | 19 | 24 | Peto-Prentice | 0.85 | 0.40 |
| CO Luminosity | 18 | 13 | Peto-Prentice | 0.44 | 0.66 |
| Log $L_{10}$ (nuclear) | 19 | 16 | Peto-Prentice | 1.87 | 0.06 |
| Log L([OIII]) | 19 | 21 | KS | 0.24 | 0.55 |
| [100, 12] | 19 | 23 | KS | 0.30 | 0.26 |
| $q$(FIR / Radio) | 19 | 23 | KS | 0.49 | <0.01 |
| $L_{8.4 GHz}$ | 19 | 24 | Peto-Prentice | 2.59 | <0.01 |

**Notes** − The statistical significance $P$ is the probability that the two samples (KSR Seyferts vs. non-KSR Seyferts) are drawn from the same parent population for the corresponding property. For the purposes of the present work, $P < 0.05$ indicates a significant difference of parent populations. Peto-Prentice (survival analysis) tests were performed for censored data (data containing upper or lower limits), but KS tests were otherwise used. There were no IRAS scans at the position of NGC 3786, and so that galaxy was removed from the comparison of $L_{12}$. Measurements of the CO luminosity, nuclear 10 μm luminosity, and [OIII] luminosity were found in the literature for most, but not all, of the Seyferts observed in the sample. The IRAS color [100, 12] is defined as the log ratio of the flux densities at those wavelengths, i.e., [100, 12] ≡ log($S_{100}$ / $S_{12}$). The tests do not include the higher redshift source Mrk 9, which was observed serendipitously.

**Table 6 KSR Position Angle Comparisons**

| Source | MERLIN / VLA (°) | Scale (″) | ΔPA (°) | VLBI (°) | Scale (mas) | ΔPA (°) | Galaxy Disk (°) | ΔPA (°) | Ref(s) |
|---|---|---|---|---|---|---|---|---|---|
| Mrk 348 | −11 | 0.1 | 24 | −15 | 2 | 28 | −45 | 58 | 1, 2, 3, 4 |
| NGC 1068 | 30 | 0.7 | 3 | 12 | 400 | 21 | 70 | 37 | 5, 6 |
| NGC 1241 | U | <0.1 | … | … | … | … | 140 | 40 | 7 |
| NGC 1320 | U | <0.2 | … | … | … | … | 138 | 48 | 7, 8 |
| NGC 2639 | 109 | 1.6 | 11 | 111 | 0.7 | 9 | 140 | 20 | 7, 9 |
| NGC 2992 | −24 | 0.3 | 54 | … | … | … | 6 | 84 | 10, 11 |
| NGC 3079 | U | <0.1 | … | 125 | 55 | 62 | 165 | 78 | 11, 12, 13 |
| NGC 3516 | 20 | 4.0 | 24 | … | … | … | 55 | 11 | 14 |
| NGC 4151 | 77 | 3.5 | 8 | 84 | 40 | 1 | 22 | 63 | 11, 15, 16, 17 |
| NGC 4235 | U | <0.06 | … | U | < 1 | … | 48 | 45 | 7, 18 |
| NGC 4051 | 81 | 0.4 | 78 | … | … | … | 135 | 48 | 19 |
| NGC 4388 | 20 | 2.0 | 13 | … | … | … | 92 | 85 | 12, 20, 21 |
| NGC 4593 | U | <0.2 | … | … | … | … | 108 | 17 | 4, 7 |
| NGC 4594 | U | <0.1 | … | … | … | … | 90 | 52 | 7, 22 |
| NGC 5347 | U | <0.3 | … | … | … | … | 130 | 52 | 7 |
| NGC 5506 | D | 3.2 | … | 73 | 29 | 67 | 95 | 45 | 7, 23 |
| NGC 5548 | U | <0.1 | … | −14 | ~4 | 1 | 110 | 55 | 7, 24 |
| NGC 5695 | U | <1.0 | … | … | … | … | 150 | 78 | 25 |
| NGC 5929 | 60 | 1.3 | 68 | … | … | … | 155 | 17 | 7, 26, 27 |

1. Neff & de Bruyn 1983
2. Ulvestad et al. 1999
3. Anton et al. 2002
4. Schmitt & Kinney 2000
5. Wilson & Ulvestad 1987
6. Gallimore et al. 1996c
7. Thean et al. 2000
8. Morganti et al. 1999
9. Wilson et al. 1998
10. Wehrle & Morris 1988
11. Carral, Turner, & Ho 1990
12. Kukula et al. 1995
13. Trotter et al. 1998
14. Miyaji et al. 1992
15. Pedlar et al. 1992
16. Pedlar et al. 1993
17. Mundell et al. 2003
18. Anderson, Ulvestad, & Ho 2004
19. Ulvestad & Wilson 1984b
20. Stone et al. 1988
21. Mundell et al. 2000
22. Hummel et al. 1984
23. Middelberg et al. 2004
24. Wrobel 2000
25. Ulvestad & Wilson 1989
26. Wilson & Keel 1989
27. Su et al. 1996

Notes. − MERLIN / VLA refers to observations arcsecond and sub-arcsecond resolution observations with either MERLIN or the VLA, or aperture synthesis measurements with comparable angular resolution (i.e., ~ 0.1″ − 1.0″). VLBI refers to mas-scale angular

resolution. In the MERLIN / VLA column, "U" means the nuclear source is unresolved, and "D" means a diffuse radio component with poorly defined PA is present.